\documentclass[12pt,preprint]{aastex}

\newcommand{\be}{\begin{equation}}
\newcommand{\ee}{\end{equation}}

\slugcomment{}

\shorttitle{Velocity Shear of the Thick Disk}
\shortauthors{Girard, et al.}

\received{2006 April 13}
\begin{document}


\title{Velocity Shear of the Thick Disk from SPM3 Proper Motions
at the South Galactic Pole
       }


\author{T. M. Girard, V. I. Korchagin\altaffilmark{1},
D. I. Casetti-Dinescu  \altaffilmark{2} and W. F. van Altena}
\affil{Yale University, Dept. of Astronomy, P.O. Box 208101, 
New Haven, CT 06520-8101
}
\email{girard@astro.yale.edu}

\author{C. E. L\'{o}pez}
\affil{Universidad de San Juan, Avenida Benav\'{i}dez 8175 Oeste, Chimbas, 
5413 San Juan, Argentina}

\author{D. G. Monet}
\affil{US Naval Obs., Flagstaff Station, P.O. Box 1149, 
Flagstaff, AZ 86002}

\altaffiltext{1}{also Institute of Physics, Rostov University, Rostov-on-Don 
344090, Russia;
Isaac Newton Institute of Chile, Rostov-on-Don Branch
}

\altaffiltext{2}{also Astronomical Institute of the Romanian Academy, 
Str. Cutitul de Argint 5, RO-75212, Bucharest 28, Romania}

\begin{abstract}

The kinematical properties of the Galactic Thick Disk are studied
using absolute proper motions from the SPM3 Catalog 
and 2MASS near-infrared photometry for a 
sample of $\sim$1200 red giants 
in the direction of the South Galactic Pole.
The photometrically-selected sample is dominated by Thick Disk stars, 
as indicated
by the number-density distribution that varies with distance from
the Galactic plane as a single-valued exponential over the range
$1<z<4$ kpc.
The inferred scale height of the Thick Disk is 0.783 $\pm$ 0.048 kpc.
The kinematics of the sample are also consistent with disk-like motion.  
The $U$-velocity component is roughly constant,
reflecting the Sun's peculiar motion,
while a considerable shear is seen in the mean rotational velocity, $V$.
The $V$-velocity profile's dependence on $z$ is linear,
with a gradient of $dV/dz = -30 \pm 3$ km s$^{-1}$ kpc$^{-1}$. 
The velocity dispersions, in both $U$ and $V$, show a lesser gradient of
about 9 $\pm$ 3 km s$^{-1}$ kpc$^{-1}$.
We demonstrate that the derived velocity and velocity-dispersion profiles
are consistent with the assumptions of dynamical
equilibrium and reasonable models of the overall Galactic potential. 

\end{abstract}

\section{Introduction}

The existence of a second, disk-like population in our Galaxy
was established on the basis of starcounts by Gilmore \& Reid (1983).
This population was referred to as a `thick disk', as its derived vertical 
scale height was greater than that of the previously known Galactic disk.
Since then, a considerable number of studies have helped
to characterize the Galaxy's Thick Disk component, in an effort to 
deduce its origin and better understand its nature. 
As a result, 
some of the mean properties of the Thick Disk are now well-established.
It consists of old ($\ga 10$ Gyr) stars of intermediate metallicities,
(-1.0 $\la$ [Fe/H] $\la$ -0.2).
These stars are kinematically hotter and possess a substantial rotational 
lag (see e.g., Majewski 1993) relative to old, Thin Disk stars. 
Most estimates of the Thick-Disk scale height are between 0.8 and 1.2 kpc,
and its local normalization is between 2\% and 8\%, although these 
two quantities appear to be somewhat anticorrelated (Siegel et al. 2002).

Among the less well-determined parameters of the Thick Disk
are its radial scale length and the magnitude of its rotational lag as
well as that of its velocity dispersions. 
Recent determinations of the radial
scale length indicate that it is larger than
that of the Thin Disk (Robin et al. 1996, Ojha 2001, Chen et al. 2001,
Larsen \& Humphreys 2003). 
There are indications that the rotational lag and the
velocity dispersions vary with distance from the
Galactic plane, (see the review by Majewski 1994). 
However, the situation is unclear regarding this matter as early
studies did not attempt
to separate the Thick Disk population from that of the Halo.
Thus, the observed variation of the lag with $z$ might be due in part 
or in whole to the mixture of the two populations. 
More modern studies 
(e.g., Majewski 1992, Chiba \& Beers 2000, 
Soubiran et al. 2003) explicitly separate
the two populations by metallicity when determining the kinematical parameters.
Unfortunately, these studies do not agree.
For instance,
Soubiran et al. (2003) find no significant kinematic gradients, 
although their study reaches only to $z=0.8$ kpc, while 
Majewski (1992), whose deep proper-motion survey
samples the Galaxy out to $z \sim 6$ kpc, 
does detect significant gradients, with that of
the rotational lag being -21 km s$^{-1}$ kpc$^{-1}$. 
Chiba \& Beers (2000)
find an even steeper gradient for the lag, -30 km s$^{-1}$ kpc$^{-1}$.
It is important to note that the Chiba \& Beers study covers a large portion of 
the sky, unlike most studies that are made only toward the Galactic poles.
It also makes use of a large sample of stars ($\sim 1200$) with known
metallicities, proper motions, radial velocities, and distance estimates.
We regard this study as a significant step forward in understanding
both the Thick Disk and the inner Halo -- its primary limitation being
that it only probes the Galaxy to $z \la 2$ kpc. 
Lastly, a very recent study (Allende Prieto et al. 2006) that uses  
SDSS photometric distances and radial velocities alone,
finds a rotational lag gradient of -16 km s$^{-1}$ kpc$^{-1}$,
between $z = 1$ and 3 kpc. 

From this brief summary, it is obvious that the kinematics of the Thick Disk
are still poorly known.
We note the following case, for example, in which
a proper interpretation of the data hinges
upon better knowledge of the Thick-Disk rotational $z$-gradient. 
Recently, Gilmore et al. (2002) (see also Wyse et al. 2006)
presented radial-velocity results of some 2000 faint F/G stars observed
in two lines of sight chosen specifically to probe Galactic 
rotation. 
They found that many of these stars, residing up to 5 kpc from the
Galactic plane, rotate at $\sim 100$ km s$^{-1}$, 
i.e., with a lag of $\sim$ -120 km s$^{-1}$. 
Their interpretation, which assumes a single-valued lag for the Thick Disk
of -35 km s$^{-1}$, is that these high-lag stars are debris from a disrupted
satellite that merged with the Galaxy in a significant accretion event. 
Alternatively, one might also explain these observations as the
natural result of a Thick Disk whose lag varies substantially with $z$,
such as has been observed by Chiba \& Beers (2000).

Finally, we note that while previous studies have attempted to
determine the variation of the Thick Disk's rotational lag and
velocity dispersions as functions of $z$, there have been no attempts
to examine the resulting velocity and dispersion profiles for self-consistency
within the context of the expected dynamical equilibrium of a disk system, 
a question first posed by Murray (1986). 
It is the purpose of the present study to determine the kinematical
properties of the Thick Disk as functions of $z$ out to $\sim$4 kpc, and
to investigate the dynamical equilibrium of this system. 
The sample used in our analysis consists of $\sim 1200$ Thick-Disk
red giants at the South Galactic Pole.
These are selected photometrically from 2MASS, and have absolute
proper motions taken from the SPM3 Catalog (Girard et al. 2004). 

In the following section, we provide details of the sample selection.
In Section 3, we describe the methods used to determine first the observed 
and then the intrinsic spatial and kinematical parameters of the Thick Disk. 
Section 4 contains our equilibrium analysis of the resulting Thick Disk
parameters, followed by a general discussion of our findings in Section 5.
Our main results are summarized in Section 6.

\section{Sample Selection}

The most recent release by the Yale/San Juan Southern Proper Motion program
is the SPM3 Catalog (Girard et al. 2004).
This catalog of absolute proper motions 
of over 10 million stars 
is complete to $V$ = 17.5
and covers an irregular area between declinations -20$\degr$ and -45$\degr$, 
excluding the Galactic plane.
We combine SPM3 proper motions with 2MASS photometry for a sample of
bright, high-latitude red giants, allowing us to examine the transverse
velocity structure of the Thick Disk.

The sample is selected from $\sim$1700 stars in the SPM3 Catalog within 
15 degrees of the South Galactic Pole (SGP), and within the region of the 
2MASS $J,K$ color-magnitude diagram shown in Figure 1.
The sloping cutoff limit at the faint end is meant to assure that 
only the most nearby (d$<$63 pc) dwarfs might contaminate the sample. 
The local density of low-mass dwarfs was tabulated by Holmberg \& 
Flynn (2000, see their Table 1), and is roughly 0.025 M$_{\odot}$/pc$^{3}$
over the color range of our sample.
Thus, within the corresponding volume of our sample cone, within 63 parsecs
for dwarf-star magnitudes, we expect on the order of several hundred such 
stars within our sample volume.
As will be seen in Section 3.1, if we calculate stellar velocities assuming
red-giant distance moduli for the sample stars, it is relatively easy to choose
a conservative velocity cutoff that cleans the sample of these kinematically
obvious nearby dwarfs.
When done, this results in the reduction of the sample size to 
$\sim 1200$ stars.

\begin{figure}
\epsscale{0.80}
\plotone{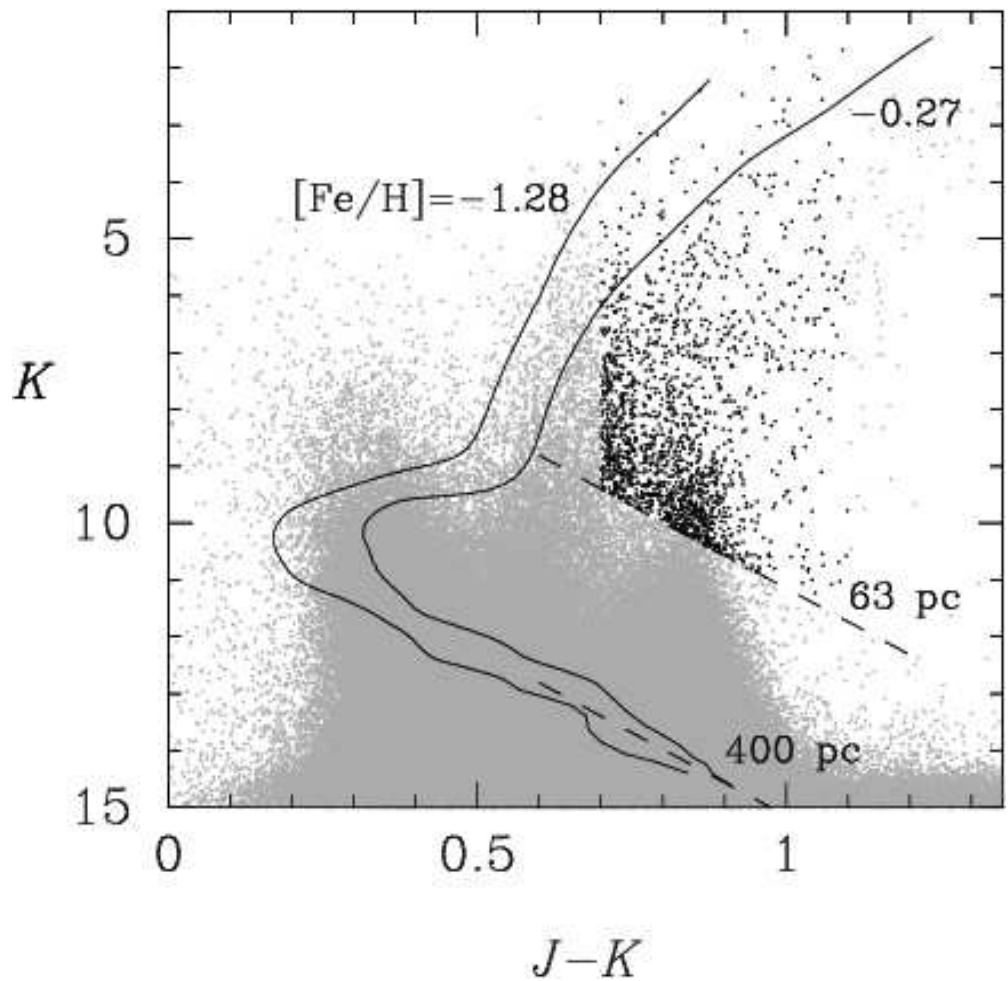}
\caption{
Color-magnitude selection of the sample, based on 2MASS $JK$ photometry.
The sample consists of stars within 15 degrees of the SGP,
with 0.70 $<$ $J-K$ $<$ 1.10, and with $K$-magnitude brighter
than the sloping limit shown.
For comparison, 5-Gyr Yale-Yonsei theoretical isochrones (Yi et al. 2003)
for two different metallicities are overlayed, placed at an arbitrary
distance of 400 pc.  The sloping cutoff used to select our sample
corresponds roughly to the main sequence at a distance of 63 pc.
}
\end{figure}

Another possible source of contamination, that from AGB-stars, is negligible.
Jackson et al. (2002) find that the
distribution of AGB-stars in the direction perpendicular
to the Galactic plane is described by an exponential function
with a vertical scale height of 0.3 kpc and a  
density of about 150 kpc$^{-3}$ in the solar neighborhood,
indicating that not more than a few AGB-stars should be in our sample.

The color range of the sample was chosen to preferentially select 
intermediate-metallicity stars (Thick Disk) as opposed to metal-poor 
ones (Halo).
As will be seen in Section 3.3, the fraction of Halo stars in our sample
can be estimated from the observed number-density profile.
This fraction turns out to be small, about 8 percent, and its presence
is included in the modeling of the sample's kinematics.

We note that the areal coverage of the sample
does not fill the entire 15-degree circular cone centered on the SGP.
The southernmost boundary of the SPM3 Catalog limits the sample to
roughly 70 percent of this volume.
Nonetheless, within this cone of irregular cross-section, 
our sample of red giants 
is expected to be volume-complete from $z$ = 0.5 to 3 kpc.

One final characteristic of our sample warrants discussion.
Monte-Carlo simulation of the observed data will be used to deduce the
sample's intrinsic kinematic properties.
This method depends on the proper-motion uncertainties for these stars
being well-determined.
Individual proper-motion uncertainties are estimated for all stars in the
SPM3 Catalog based on proper-motion differences between different image
systems and exposures.
As is often the case, these internal uncertainty values may underestimate
the true uncertainties, on average.
In order to ``calibrate'' the SPM3 proper-motion uncertainty estimates,
we select all stars in the SPM3 Catalog that are also Hipparcos stars and
examine the differences in the SPM3 and Hipparcos proper-motion 
determinations.
The number of such stars is 10,900 and the dispersion of the differences 
is 3.75 mas yr$^{-1}$.
The Hipparcos measuring errors are approximately 1 mas yr$^{-1}$, thus,
the mean SPM3 proper-motion error for these stars is 3.61 mas yr$^{-1}$.
The trimmed mean uncertainty estimates from the SPM3, for these Hipparcos
stars, is 3.12 mas yr$^{-1}$.
Therefore, adjustment by a factor of 1.16 is in order and should be applied to 
the SPM3 internal uncertainty estimates for these bright, well-measured stars.
In fact, our SGP sample stars are among the best-measured in the SPM3 Catalog,
being slightly fainter than the Hipparcos stars and at the optimal magnitude
range for the SPM plate material.
The trimmed mean, SPM3-estimated proper-motion uncertainty for the SGP 
sample is 2.58 mas yr$^{-1}$.
We adopt 1.16 times this value, 3.0 mas yr$^{-1}$, as the actual proper-motion
uncertainty for our SGP red-giant sample.

\section{Proper-Motion/Photometry Analysis}

Our goal is to derive the tangential velocity distribution of the Thick
Disk as a function of distance from the Galactic plane, $z$, based on our
proper-motion sample.
Details of how this is accomplished are given below, but the general
procedure is as follows:
Individual stellar distances are estimated photometrically, using
2MASS $K$ apparent magnitudes and an adopted absolute-magnitude
calibration.
At a given distance, a star's ($U,V$) velocity components are then 
simply computed from its measured proper motion, ($\mu_U, \mu_V$).
The average velocity and velocity dispersion of the sample, as a function of
$z$, is then determined.

Uncertainties in the derived distances will affect the sample's ``observed'' 
density, velocity, and velocity-dispersion profiles, altering them from 
their actual, intrinsic forms.
There are several distinct mechanisms associated with the sample selection
and distance estimation that combine to produce a bias 
in the observed profiles.
Chief among these is the convolution of the distance errors with the
sharply rising frequency distribution of stars as a function of $z$.
It is essential that compensation be made for all these effects, as well as
those caused by uncertainties in the measured proper motions.
We do so by modeling the effects on simulated data, as described in
Section 3.2.
The conversion from measured quantities (apparent magnitudes, proper
motions) to stellar parameters (distance, velocity) is described first.

\subsection{Observed Sample}

Our sample has been cut from a portion of the $J,K$ color-magnitude
diagram to preferentially select red giants at distances of 0.5 to 4
kpc. The sample is dominated by Thick-Disk, 
intermediate-metallicity giants, so the expected absolute magnitudes of
these stars can be estimated as a function of color.
Yale isochrones (Yi et al. 2003)
have been used, and convolved with an adopted Thick-Disk metallicity
distribution to yield absolute $M_K$ magnitude distributions as a function
of $J-K$ color bin.
The assumed metallicity distribution is taken from Wyse \& Gilmore (1995), 
(the intermediate component of their Figure 16).
The resulting absolute-magnitude distributions that we adopt are shown
in Figure 2.

\begin{figure}
\epsscale{0.80}
\plotone{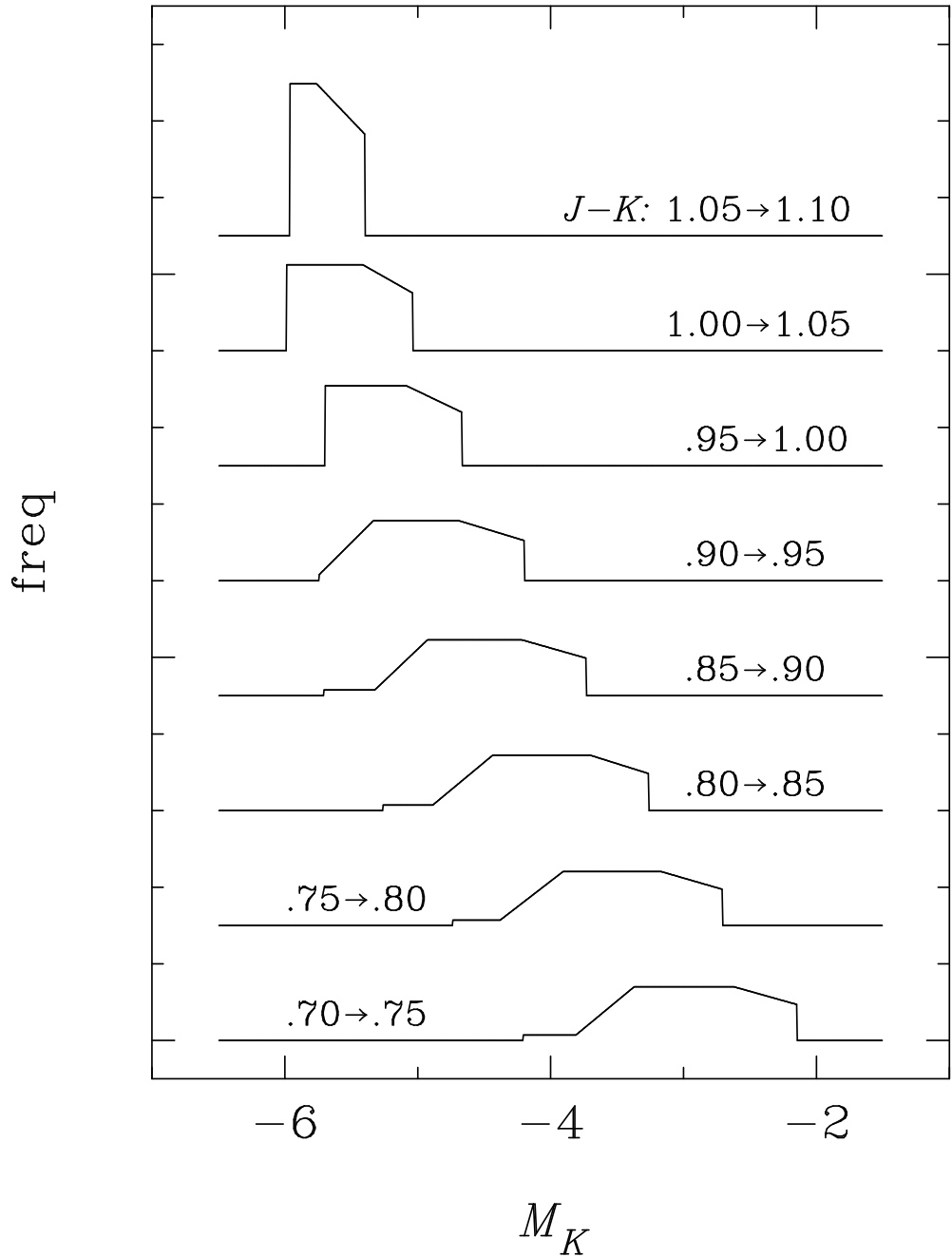}
\caption{
Assumed red-giant absolute magnitude distribution, by color bin, used in
the photometric distance determinations.
These distributions were constructed by combining the Thick-Disk
metallicity distribution measured by Wyse \& Gilmore (1995) with
Yale-Yonsei theoretical isochrones, (Yi et al. 2003). 
}
\end{figure}

For a particular star with known apparent magnitude $K$, the range in 
absolute magnitude corresponds to a range in derived distance.
We distribute each star over a range of distance by artificially
partitioning the star into 100 subunits and randomly drawing a value of 
$M_K$ for each subunit, utilizing the absolute-magnitude distribution
as a probability distribution.

Velocity components are derived for each subunit, thus distributing the star
in velocity space as well.
Note that the uncertainties in the extinction-corrected values of $K$
and $J-K$ are negligible relative to the width of the absolute-magnitude
distributions.
In this manner, the absolute proper motions as a function of apparent magnitude,
shown in Figure 3, are converted to velocities as a function of $z$,
shown in Figure 4.
Obviously, a substantial fraction of the derived velocities are not
physically meaningful.
Nearby dwarfs misinterpreted as giants are assumed too distant
and erroneously given large velocities.
The large increase in proper-motion dispersion beyond $K=9$ in Figure 3
is precisely due to the incursion of nearby dwarfs into the sample.
The velocities calculated for these stars, under the assumption that they
are giants, fall well outside the trend of the Thick Disk giants seen in
Figure 4, often beyond the plotted limits of the velocity axis, in fact.
To minimize this source of contamination, we cut those (subunits of) stars
which have velocities greater than the local escape velocity, i.e.,
in excess of 550 km s$^{-1}$ relative to an assumed LSR velocity 
of $V$ = -220 km s$^{-1}$.
This trimming effectively decreases the size of our SGP sample from 1700 
stars to roughly 1200 stars.

\begin{figure}
\epsscale{0.80}
\plotone{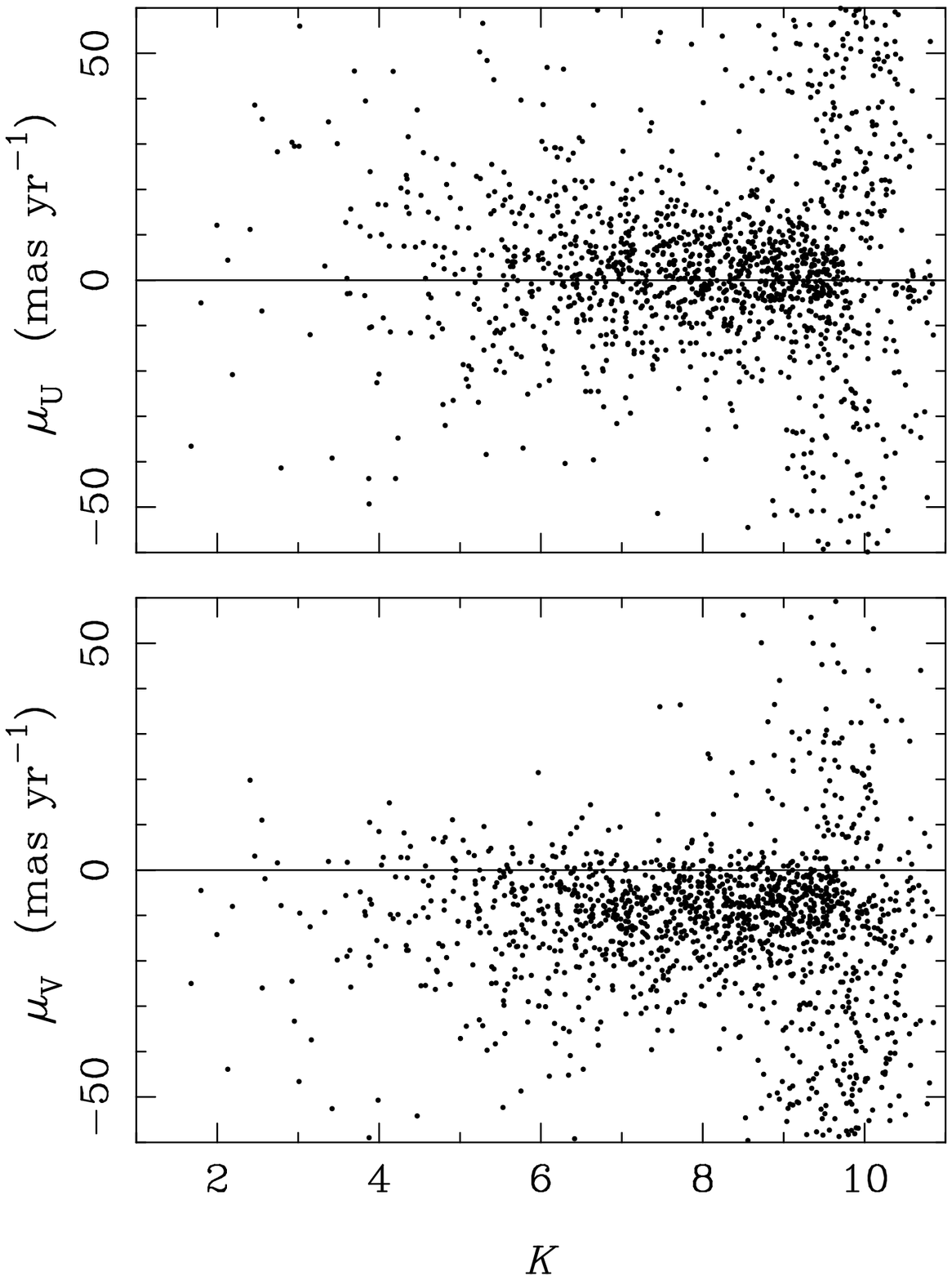}
\caption{
Absolute proper motions as a function of magnitude for the photometrically
selected red-giant sample.
$U$ is positive outward from the Galactic center, and $V$ is in the direction
of Galactic rotation.
}
\end{figure}

\begin{figure}
\epsscale{0.80}
\plotone{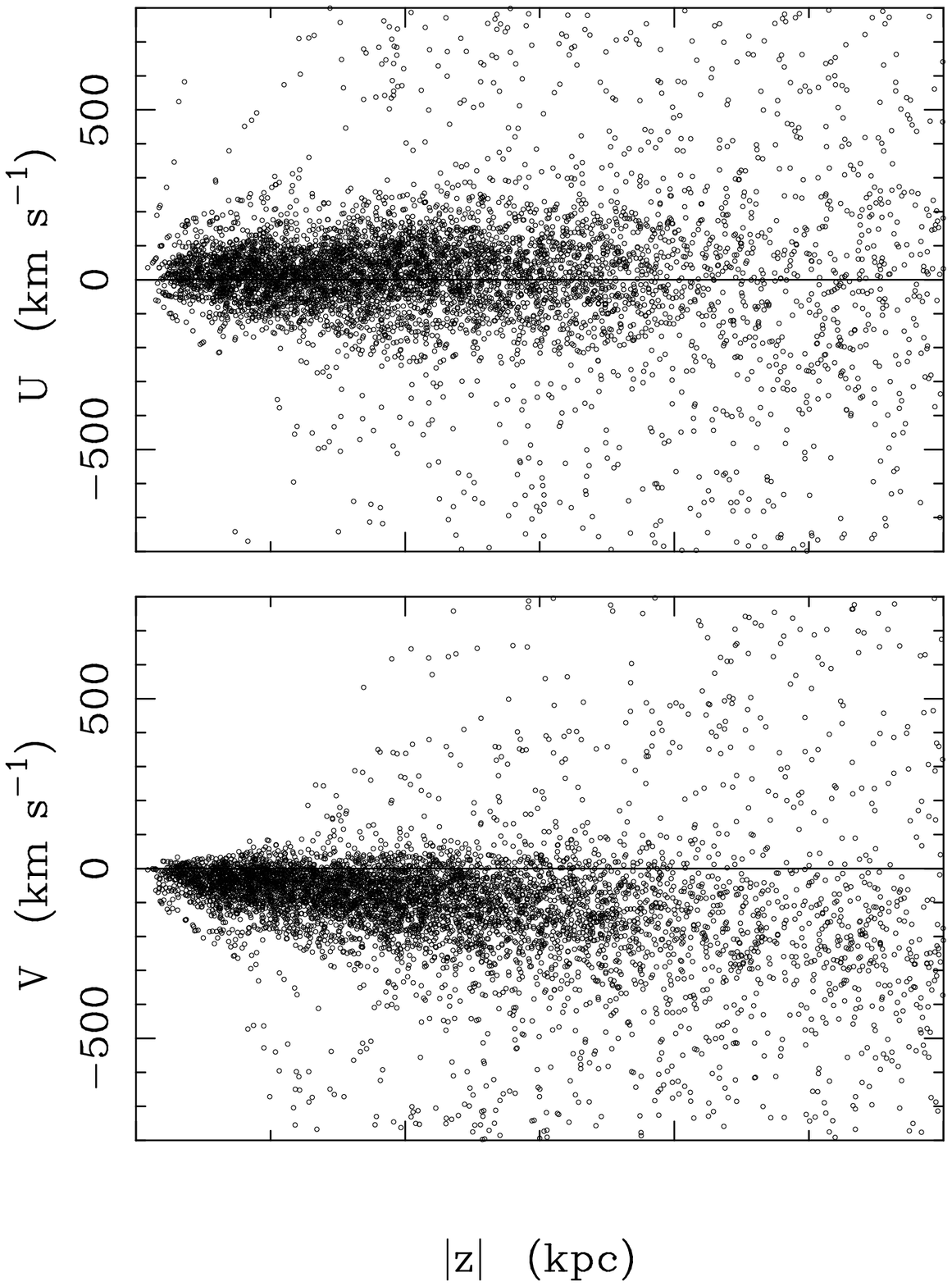}
\caption{
Tangential velocities as a function of distance from the Galactic plane, $z$.
Each star is assigned a distribution of 100 absolute magnitudes, randomly
drawn from the luminosity functions given in Figure 2.  Each point in
the plot represents one of the 100 such ``subunits'' per star.
}
\end{figure}

An alternative to velocity trimming of the sample would be to use the
reduced proper motion diagram to separate giants and dwarfs.
The reduced proper motion is a kinematic estimate of absolute magnitude
and is given by
\[ H_K = K + 5 log|\mu| + 5, \]
where $|\mu|$ is the total magnitude of the star's proper motion, 
in arcsec yr$^{-1}$.
In Figure 5, the reduced proper motion in $K$-band is plotted versus $J-K$
for our SGP sample.
The result of the subunit velocity trimming we have adopted is illustrated by
the different symbols in the figure.
Those stars for which 95\% or more of their subunits pass the velocity 
trimming are shown as solid symbols.
These, presumably, are the desired giants.
Stars for which 5\% or fewer of their subunits pass the velocity
trimming are shown as open circles.
These are more nearby dwarfs.
A smaller number of stars, shown as crosses, 
have between 5\% and 95\% of their subunits
that satisfy the velocity cut.
These lie almost exclusively in the
less populated region between giants and dwarfs in the diagram.
Thus, the subunit velocity trimming produces a giant/dwarf segregation
that is virtually identical to that which would have resulted from a 
simple cut by reduced proper motion.
The subunit velocity trimming provides a means of partially including
borderline stars in the sample, but only those subunits for which the
randomly selected absolute magnitudes lead to reasonable velocities.

\begin{figure}
\epsscale{0.80}
\plotone{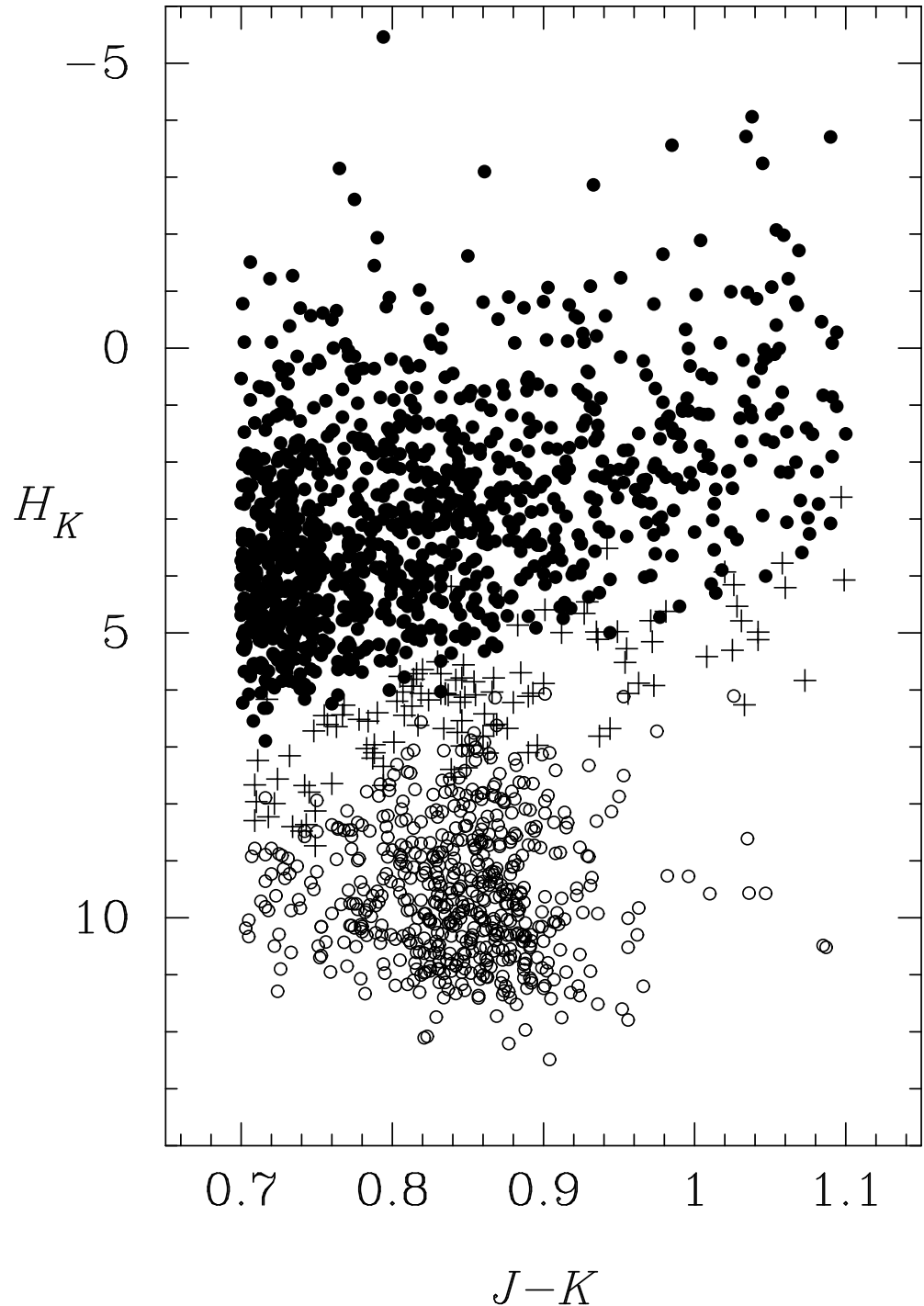}
\caption{
Reduced proper motion diagram for the SGP sample.
Symbols represent delineation in the fraction of each star's subunits
that survive the velocity trimming, i.e. total velocity relative to that
of the LSR less than 550 km s$^{-1}$.
Solid dots are stars with 95\% or more subunits passing the trimming.
Open circles are stars with 5\% or fewer of their subuints passing the cut.
Crosses represent the remaining stars whose subunits survive the velocity
trimming with a frequency that falls between these two values.
}
\end{figure}

After velocity trimming, the distributions of star density, velocity,
and velocity-dispersion are computed as functions of $z$.
The resulting distributions are shown in Figure 6.
These profiles are constructed by sorting the subunits in $z$ and binning
them, 100 subunits per bin.
Thus, each point in Figure 6 has the weight of one star, although in general
a point will have contributions from numerous stars, effectively smoothing
the data.
The average (median-based) velocity and velocity dispersion of each bin 
is estimated
using the probability-plot method, (Hamaker 1978), using the inner 80 percent
of each bin's velocity distribution.
This leads to a more robust dispersion estimate, one that is less
influenced by outliers - whether they be actual mismeasures or a small 
fraction of contaminating stars from a kinematically hotter population, 
i.e., Halo stars.

\begin{figure}
\epsscale{0.58}
\plotone{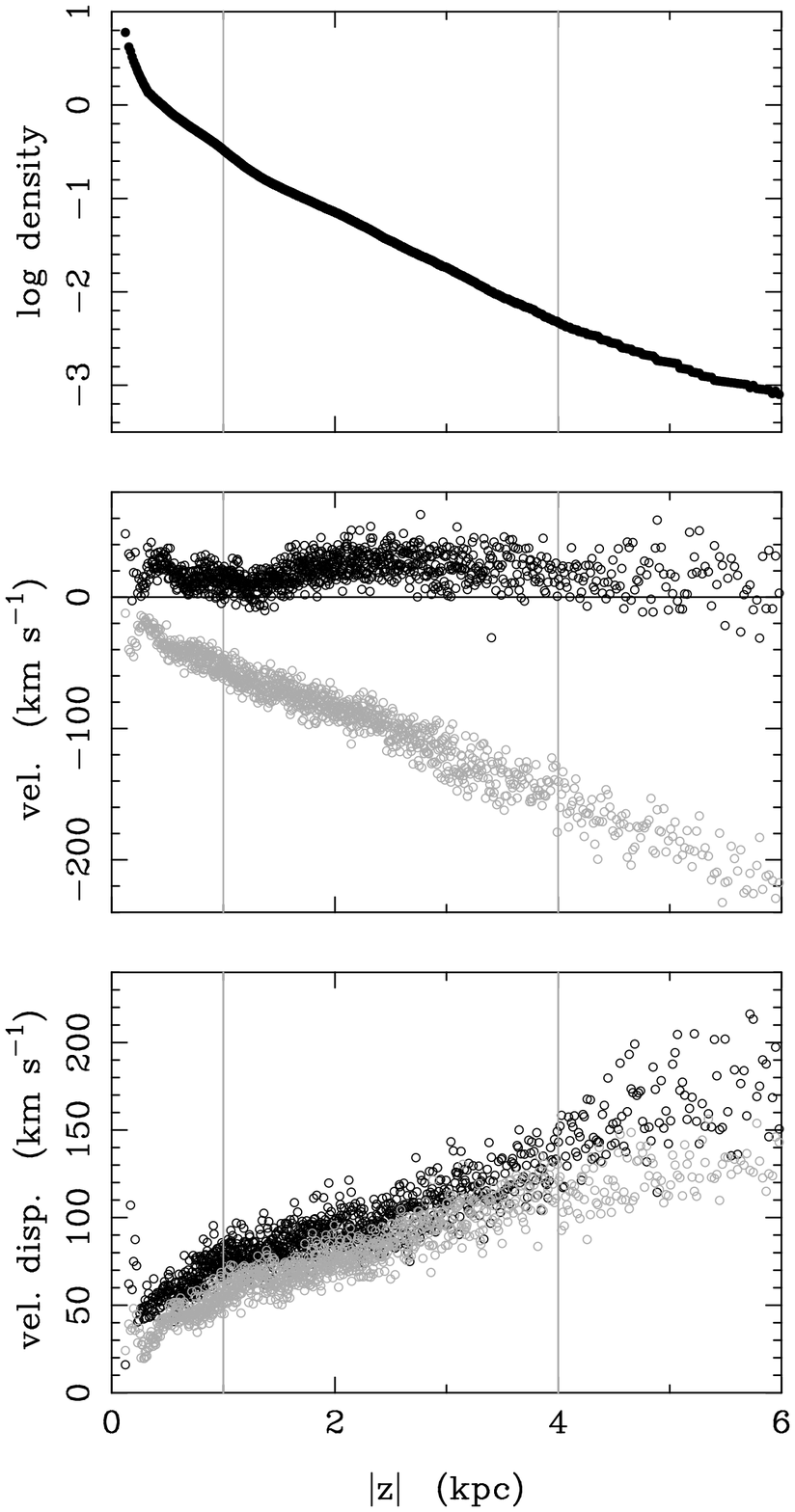}
\caption{
Stellar density, velocity, and velocity-dispersion as a function of $z$
for the observed sample of $\sim$1200 red giants.
Star ``subunits'' are sorted in $z$ and then binned in groups of 100 to
calculate the total density, mean velocity, and dispersion in each bin.
In the lower two panels, dark symbols are for the $U$ velocities and lighter
symbols are for the $V$ velocities.
}
\end{figure}

The observed log-density profile in Figure 6 indicates
a nearly linear trend in the range 1 to 4 kpc.
Based on the imposed faint-limit cutoff of our sample and the bright
limit of the SPM3 Catalog, combined with the assumed 
absolute-magnitude distributions, we estimate that our red-giant sample
is expected to be complete from approximately $z$ = 0.5 to 3 kpc.
There are appreciable deviations from otherwise smooth trends in the
range of 0.5 to 1 kpc in all of the observed profiles.
These possibly show the presence of Thin Disk giants over this range
and/or small-number fluctuations, keeping in mind that the volume
density of stars is proportional to the counts but inversely proportional
to the square of the distance for our sampling cone.
On the other hand, in the range $z$ = 3 to 4 kpc, while some incompleteness
is expected, the various observed profiles continue their trends
across this region.
Thus, we choose to concentrate our analysis on the range $z$ = 1 to 4 kpc.
It is over this range that we will attempt to parameterize the underlying
density, velocity and dispersion profiles of our sample.

\subsection{Simulated Samples}

As stated earlier, a correction must be made to the observed profiles
to compensate for systematic effects caused by measurement uncertainties.
The strategy we employ is to characterize the distance and proper-motion
uncertainties in detail and apply them to simulated data sets, in a Monte
Carlo fashion.
The generated samples are assigned artificial errors in distance (via
the absolute magnitudes) and in proper motion, consistent with the
uncertainty distributions of the actual SGP sample.
The simulated data sets are then used to determine the
transformation between the intrinsic and ``observed'' density, velocity, and
velocity-dispersion profiles.
This transformation, when inverted, can then be used to determine the
intrinsic parameters of our SGP sample from its observed profiles shown
in Figure 6.
Once the intrinsic parameters are determined, their uncertainties can be
estimated in a standard Monte Carlo approach.
That is, a number of simulations
with these same intrinsic parameters is run, and 
the ``observed'' profiles are
generated and then used to estimate the input intrinsic parameters.
The distribution of derived intrinsic parameters about the known input
values provides an estimate of their uncertainties.

The model we construct for our simulations consists of two components,
a planar component having an exponential density distribution in $z$ which we
identify with the Thick Disk, and a spherical component with a power-law
density distribution in Galactocentric radius and which we identify
with the Inner Halo.
The spatio-kinematic characteristics of the Halo component are assumed
known, and will be taken from the literature, 
whereas those of the Thick Disk component are the subject of
this investigation and are to be determined.

The power-law index for the Halo density is assumed to be -3.5, 
(Zinn 1985).
A somewhat less negative index might be appropriate for the inner Halo, but
at the adopted Galactocentric
distance of the sun, assumed to be 8 kpc, the Halo density changes very
little over the $z$ = 1 to 4 kpc region of interest for any reasonable choice
of index.
Thus, the precise value of the power-law index is not critical.
The velocity and velocity dispersion for the Halo component are assumed
to be independent of $z$.
Values are taken from Table 2 of Chiba and Beers (2000), using
mean parameters of their more distant samples with [Fe/H] $\lesssim$ -1.5.
The adopted Halo velocity parameters are:
$U_{H}$ = 16 km s$^{-1}$, 
$V_{H}$ = -180 km s$^{-1}$, 
$\sigma_{U H}$ = 150 km s$^{-1}$,
$\sigma_{V H}$ = 120 km s$^{-1}$.
The only free parameter associated with the Halo component 
is the fraction of Halo stars, $f_H$, 
which is to be determined for our SGP sample.

The simulated Thick Disk component is parameterized as follows.
The number-density distribution is assumed a pure exponential, with scale
height $h_{z_{thick}}$.
The velocity and velocity dispersion in $U$ and $V$ are expressed as linear
functions of $z$.
It is not that we have theoretical reasons to suspect linear relations,
but simply that the observed velocity and dispersion profiles show an
apparent linear trend over the region of interest and we do not feel that
the data would allow us to deduce terms higher than first-order.
Thus, it is assumed that the intrinsic velocity and dispersion 
profiles can be adequately described by linear functions

\[ U = U_o + U'|z|, \]
\[ V = V_o + V'|z|, \]
\[ \sigma_U = \sigma_{Uo} + \sigma'_U|z|, \]
\[ \sigma_V = \sigma_{Vo} + \sigma'_V|z|. \]

A simulation is characterized by the ten parameters 
$f_H$, $h_{z_{thick}}$, $U_o$, $U'$, $V_o$, $V'$, 
$\sigma_{Uo}$, $\sigma'_U$, $\sigma_{Vo}$, $\sigma'_V$.
The desired number of Halo giants is generated, falling within a 15-degree
cone at the SGP and randomly selected from the power-law distribution in
Galactocentric radius, $r$.
Thick Disk giants are similarly generated within the cone and distributed
with an exponential falloff in $z$, with scale height $h_{z_{thick}}$.
Each star is given a $J-K$ color drawn from a distribution that follows
the colors of the actual SGP sample.
Based on the color, an absolute magnitude $M_K$ is assigned, drawn from
the appropriate magnitude distribution of Figure 2.
A slightly different set of low-metallicity $M_K$ distributions is employed
for the Halo stars.
Once the distance and absolute magnitude are assigned, the apparent magnitude
is calculated and checked to see if it falls above the sloping 
apparent-magnitude cutoff used to define the SGP sample.
If so, the star is retained and given randomly selected $U$ and $V$ 
velocities, drawn from the appropriate velocity-$z$ relations for that 
stellar component.
Gaussian distributions about the mean velocity trends are assumed, with
dispersions given as a function of $z$ as specified by the input parameters.
The combination of $U$ and $V$ with $z$, and Galactic latitude, specify
the star's precise proper motion, $(\mu_U,\mu_V)$.
The ``observed'' proper motion is generated by adding a random deviate
chosen from a Gaussian distribution whose standard deviation is 
3.0 mas yr$^{-1}$, the measuring uncertainty determined for our SGP sample.
Thick Disk and Halo giants are accumulated in the desired ratio, based on
$f_H$, until the sample consists of 1160 stars, the approximate size of
the velocity-trimmed SGP sample.

After the simulated sample is generated, it is passed through the same
reduction procedure that was applied to the actual SGP sample.
Note that absolute magnitudes are derived based only on the 
intermediate-metallicity
curves of Figure 2, as at this point the distinction between Thick Disk
and Halo stars is intentionally discarded.
This is not too crude of an approximation as the magnitude distributions for
the low- and intermediate-metallicity giants do not vary greatly, and the
fraction of Halo stars turns out to be small.
More importantly, any systematic effect caused by this approximation will
also be present in the actual SGP sample's reduction and should, therefore,
be mimicked here.
Density, velocity, and dispersion profiles are constructed for each 
simulation in the same way as was done for the SGP sample.
A representative example is shown in Figure 7, for a simulation with
input parameters very near the final best-fit values for the SGP sample.

\begin{figure}
\epsscale{0.58}
\plotone{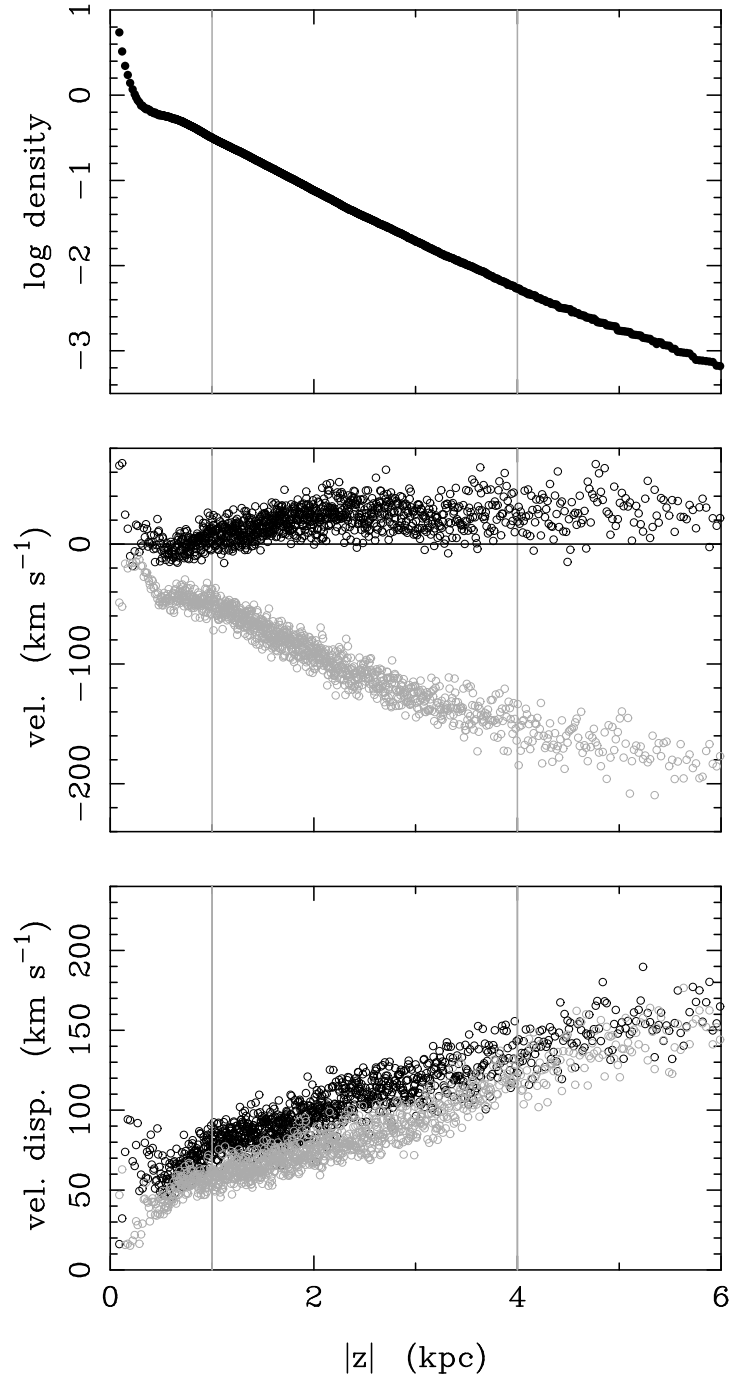}
\caption{
Stellar density, velocity, and velocity-dispersion as a function of $z$
for a single simulated sample.  The construction of the distributions and
the plotting symbols are the same as used for the observed sample, shown in
Figure 6.  The spatio-kinematic input parameters of this simulation are
very near those of the best-fit determination given in Section 3.3
}
\end{figure}

\subsection{Intrinsic Parameter Estimation}

The primary purpose of the simulations is to provide a means of determining
the relationship between the ten input model parameters and the simulated
``observed'' density, velocity, and dispersion profiles.
The various parameters are evaluated as follows.

The value of the Halo fraction influences all of the profiles in Figure 7.
However, the density profile provides the most direct measure of this
parameter.
The slope and curvature of the log-density plot depends
on the combination of $h_{z_{thick}}$ and $f_H$.
We characterize the output profile by fitting a quadratic
polynomial to the log-density as a function of $z$, from $z$ = 1 to 4 kpc,
\[ log(\rho) = a_0 + a_1 |z| +a_2 |z|^2. \]
The values of $a_1$ and $a_2$ will be functions of the input parameters
$h_{z_{thick}}$ and $f_H$.
This functional dependence is determined by generating simulations
on a 2-d grid that spans a suitable range of the input parameters,
(i.e., $ 0.0 < f_H < 0.4,  0.6 < h_{z_{thick}} < 0.9 $ kpc).
At each grid point a series of 25 simulations was performed and means of
the best-fit values of $a_0$, $a_1$, and $a_2$ determined.
Linear fits of $a_1$ and $a_2$ as functions of the input parameters
were then found by least-squares,
\[ a_1 = -3.74 + 2.87 h_{z_{thick}} - 0.15 f_H  \]
\[ a_2 = 0.30 - 0.367 h_{z_{thick}} + 0.18 f_H \]
where $h_{z_{thick}}$ is in kpc.
The above can be inverted to provide the desired means of estimating the
intrinsic parameters, $h_{z_{thick}}$ and $f_H$, given a measurement of the
observed density-profile parameters, $a_1$ and $a_2$.

For our SGP sample, the least-squares-determined values of $a_1$ and $a_2$
are -1.51 and 0.027, respectively.
Substituting these into the above relations and solving yields
$f_H = 0.08$ and $h_{z_{thick}} = 0.783$ kpc.
Uncertainty estimates for these and the other intrinsic parameters will
be presented in the following subsection.

With the Halo fraction determined, and found to be relatively low, 
we now proceed to modeling of the $U$ and $V$ velocity profiles.
The two components are treated separately 
and are assumed to be independent of one another.
Again, simulations are carried out over a grid of values, in this
case two grids, varying $U_o$ and $U'$ over an appropriate range of 
values for one grid, and varying $V_o$ and $V'$ in the other.
Least-squares fits of the resulting $U$ and $V$ profiles are made, this
time using linear functions of $z$, fitting over the same 1 to 4 kpc region.
This yields a grid of ``observed'' values of intercept and slope for the
simulated profiles.
The mapping from intrinsic velocity-profile parameters to observed ones
is made by least-squares fitting.
The resulting relationships, in the two components, are
\[ U_o = -0.696 + 1.18 \hat{U}_o - 0.373 \hat{U}' \]
\[ U' = -1.28 + 1.20 \hat{U}' \]
\[ V_o = -10.7 + 1.15 \hat{V}_o - 0.316 \hat{V}' \]
\[ V' = 8.46 + 1.21 \hat{V}' \]
where parameters marked with a \^{ } indicate observed quantities as 
opposed to intrinsic ones.

The observed profile parameters for the actual SGP sample are
$\hat{U}_o$ = 9.63 km s$^{-1}$,
$\hat{U}'$ = 5.12 km s$^{-1}$ kpc$^{-1}$,
$\hat{V}_o$ = -21.8 km s$^{-1}$,
$\hat{V}'$ = -32.1 km s$^{-1}$ kpc$^{-1}$.
From the above relations, these correspond to best-estimate values for
the sample's intrinsic velocity parameters of
$U_o$ = 8.7 km s$^{-1}$,
$U'$ =  4.8 km s$^{-1}$ kpc$^{-1}$,
$V_o$ = -25.8 km s$^{-1}$,
$V'$ = -30.3 km s$^{-1}$ kpc$^{-1}$.

A similar treatment of the velocity dispersions is made.
Linear descriptions of the observed and intrinsic dispersion profiles
are used.
As will be seen in Section 4.2, a strictly linear dependence on $z$ of both
the velocity and velocity-dispersion profiles is not consistent with
the assumed condition of equilibrium for our sample, given a reasonable
distribution for the underlying potential.
However, given the quality of the data over the applicable range
in $z$, descriptions of the profiles that include higher than first-order
terms are not warranted.
With this in mind, the derived relationship between observed and intrinsic 
profile linear parameters is
\[ \sigma_{Uo} = 26.1 + 1.16 \hat{\sigma}_{Uo} - 0.853 \hat{\sigma}'_U \]
\[ \sigma'_U = -22.2 + 1.62 \hat{\sigma}'_U \]
\[ \sigma_{Vo} = 26.1 + 1.16 \hat{\sigma}_{Vo} - 0.976 \hat{\sigma}'_V \]
\[ \sigma'_V = -24.6 + 1.75 \hat{\sigma}'_V \]
where again a variable with a \^{ } indicates an observed quantity.

The observed dispersion profile parameters of the SGP sample are
$\hat{\sigma}_{Uo}$ = 55.7 km s$^{-1}$,
$\hat{\sigma}'_U$ = 18.3 km s$^{-1}$ kpc$^{-1}$,
$\hat{\sigma}_{Vo}$ = 36.5 km s$^{-1}$,
$\hat{\sigma}'_V$ = 20.0 km s$^{-1}$ kpc$^{-1}$.
These correspond to the following best-estimate values for the sample's
intrinsic velocity-dispersion parameters:
$\sigma_{Uo}$ = 74.8 km s$^{-1}$,
$\sigma'_U$ = 7.5 km s$^{-1}$ kpc$^{-1}$,
$\sigma_{Vo}$ = 48.7 km s$^{-1}$,
$\sigma'_V$ = 10.5 km s$^{-1}$ kpc$^{-1}$.

We now estimate the uncertainties in all of the derived intrinsic parameters.


\subsection{Parameter Uncertainties}

Uncertainties in the derived values of the intrinsic Thick Disk parameters
are estimated using Monte Carlo techniques.
A series of 100 simulations is performed, using our best-estimate values
for the intrinsic parameters.
Each simulated data set is then reduced through the same series of steps 
as was the actual SGP sample.
That is, every star is represented by 100 subunits, each with its own
photometric distance estimate; the subunits are binned by $z$; density,
velocity and dispersion are calculated for each bin, the latter two using
probability-plot methods; the resulting ``observed'' distributions are fit 
over the range $1 < z < 4$ kpc, using quadratic (density) and linear
(velocity and dispersion) functions of $z$; then these profile fitting
parameters are used to determine estimates of the ten intrinsic parameters,
as just described.
The distribution of the derived values for each intrinsic parameter
provides an estimate of the uncertainty in that parameter.

The resulting distributions are found to be roughly normal, 
for all ten intrinsic parameters.
Adopting the standard deviation of each parameter's distribution as a
one-sigma uncertainty in its determination leads to the following final
estimates for the intrinsic Thick Disk parameters:

\indent
$f_H = 0.080 \pm 0.056$  \\
\indent
$h_{z_{thick}} = 0.783 \pm 0.048$ kpc  \\
\indent
$U|_{z=2.2} = 19.1 \pm 2.7 $ km s$^{-1}$,
$U' =  4.8 \pm 2.9 $ km s$^{-1}$ kpc$^{-1}$  \\
\indent
$V|_{z=2.2} = -91.6 \pm 1.9 $ km s$^{-1}$,
$V' = -30.3 \pm 3.2 $ km s$^{-1}$ kpc$^{-1}$  \\
\indent
$\sigma_{U}|_{z=2.2} = 91.1 \pm 2.9 $ km s$^{-1}$,
$\sigma'_U = 7.5 \pm 3.1 $ km s$^{-1}$ kpc$^{-1}$  \\
\indent
$\sigma_{V}|_{z=2.2} = 71.5 \pm 1.9 $ km s$^{-1}$,
$\sigma'_V = 10.5 \pm 3.3 $ km s$^{-1}$ kpc$^{-1}$.

In the above, the constants in the velocity and velocity-dispersion
relations are evaluated at $z$=2.2 kpc, instead of $z$=0.
This is the effective mean value of $z$ for the sample over the
fitting range from 1 to 4 kpc.
Shifting the intercepts in this manner eliminates the correlation between the
constant term and slope for these linear fits, simplifying the expression
of the uncertainties in the two terms.
For example, the linear description of $U(z)$ becomes
$U = U|_{z=2.2} + U'(z-2.2)$,
where $z$ is in kpc, with similar expressions for $V$ and for both 
components of the dispersion.
Once again, we note that the choice of a linear description, in particular 
for the dispersion profiles, does not indicate that an intrinsically linear
functional form is expected, merely that the data do not allow for
the determination of higher-order terms.
The linear descriptions given above are valid over
the range $z$ = 1 to 4 kpc.

The derived Thick-Disk scale height, velocity profiles, and 
velocity-dispersion profiles will be compared to results from other
studies, and discussed further, in Section 5.
First, though, we analyze our results in the context of a relatively simple
model of the Galaxy, in an effort to demonstrate that the derived
intrinsic profiles are self-consistent, and possibly to constrain other
fundamental Galactic parameters.

\section{Equilibrium Model Comparison}

Under the assumption of dynamical equilibrium, the Thick Disk velocity
and velocity dispersion as functions of $z$ are governed by the
Galactic gravitational potential in which the Thick Disk resides.
Are the measured velocity and dispersion profiles consistent with a condition
of equilibrium and reasonable models for the Galactic potential?
Can the profiles be used to constrain models 
of the Galaxy's potential?

To address these questions, we derive expressions for the transverse 
velocity and velocity dispersion of a stellar Thick Disk in the presence of a 
parameterized model for the Galactic potential.
The resulting relation between velocity and dispersion profiles is then
used to investigate allowed values for the Galaxy-model parameters.

\subsection{Formulation}

In order to derive an expression for the rotational velocity,
$v_{\Theta}(R,z)$, of a relaxed
population of Thick Disk stars in equilibrium within the gravitational potential
of the Galaxy, $\Phi_{tot}(R,z)$,
we start with the Jeans equation:
\be
    {1 \over \rho_i} {\partial (\rho_i \overline {v_R^2}) \over \partial R} +
    {1 \over \rho_i} {\partial (\rho_i \overline {v_R v_z}) \over\partial z} +
    {\overline {v_R^2} - \overline {v_{\Theta}^2} \over R} +
    {\partial \Phi_{tot} \over \partial R} = 0
    \label{eq1}
\ee
Here $R$, $\Theta$ and $z$ are galactocentric cylindrical coordinates,
$\rho_i(R,z)$ is the volume density of a relaxed population of Thick Disk
stars, and $\Phi_{tot}(R,z)$ is the total gravitational potential of the Galaxy.
Defining velocity dispersions of the Thick Disk stars as
$$
     \overline {v_{\Theta}^2} = \sigma_{\Theta}^2 + {\overline {v_{\Theta}}}^2,
$$
$$
     \overline {v_R^2} = \sigma_R^2
$$ 
one can express the rotational velocity $\overline {v_{\Theta}}$ from 
Equation (1) as:
\be
{\overline {v_{\Theta}}}^2 = \sigma_R^2 \Big( {R \over \rho_i \sigma_R^2}
{\partial (\rho_i \sigma_R^2) \over \partial R} + {R \over \rho_i \sigma_R^2}
{\partial (\rho_i \overline {v_R v_z}) \over \partial z} + 1 - {\sigma_{\Theta}^2 \over
\sigma_R^2 } \Big) 
+ R {\partial \Phi_{tot} \over \partial R}
\label{eq2}
\ee 

Evaluation of the first term on the right side of Equation (2), 
the partial derivative
with respect to $R$, requires knowledge of the $R$ dependence of both the
Thick-Disk number density and the Thick-Disk velocity dispersion.
We assume that the density distribution can be represented by exponential
distributions in the radial and vertical directions,
$$
  \rho_i (R,z) = \rho_0 exp (-(z/h_{z_{thick}} + R/h_{R_{thick}}) \mbox {,}
$$
where $h_{R_{thick}}$ and $h_{z_{thick}}$ are the respective scale length and
scale height.
Star counts from the APS Catalog (Larsen \& Humphreys 2003) give an
estimate for the radial scale length of the Thick-Disk number density of
$4.7 \pm 0.2$ kpc.
Analysis of SDSS data
(Juri\'{c} et al 2005) gives a value of $\sim$ 3.5 kpc.
We shall explore values ranging from 3.5 to 5 kpc in our modeling.

There have been no attempts to estimate the radial scale length
of the Thick-Disk velocity-dispersion profile.
If the Thick Disk were self-gravitating, one would expect the square
of the {\it vertical} velocity dispersion to be proportional to the density. 
If one further assumes that, as with the Thin Disk, the shape of
the velocity ellipsoid of
the Thick Disk is not varying with $R$ at the solar circle,
then ${\sigma_R}^2 \varpropto {\sigma_z}^2$.
(For example, this is what one would expect if the scattering of
individual Thick Disk stars is an isotropic process.)
Consequently, this implies that
${\sigma_R}^2$ will also follow the Thick-Disk density distribution.
Under these assumptions, the first term within the parentheses in
Equation (2) 
would reduce to
\begin{mathletters}
\begin{eqnarray}
-{ 2 R \over h_{R_{thick}}} \equiv - \Upsilon_a R.
\end{eqnarray}
As will be discussed shortly, the total mass of the Thin Disk
substantially exceeds
that of the Thick Disk and, likewise, gradients in the potential in which the
Thick Disk resides are expected to be dominated by the Thin-Disk mass 
distribution.
Therefore, it may be more reasonable to assume that the vertical velocity
dispersion of the Thick Disk stars will follow the mass distribution of
the Thin Disk.
Again, under the assumption of a non-varying shape of the Thick-Disk
velocity ellipsoid, this implies the following form for the first term
in Equation (2),
\begin{eqnarray}
-{ \Big( { {1 \over h_{R_{thin}}} + {1 \over h_{R_{thick}}} } \Big) R }
\equiv - \Upsilon_b R,
\end{eqnarray}
\end{mathletters}

where $h_{R_{thin}}$ is the radial scale length of the Thin Disk.

These two alternative expressions for the factor $\Upsilon$,
Equations (3a) and (3b),
are reasonable limiting
cases for a generic thick disk.
They represent the totally self-gravitating
and the totally non-self-gravitating approximations, {\it i.e.} the latter
assumes total dominance of an imbedded thin disk.
The true nature of the Milky Way Thick Disk presumably lies 
somewhere between these limits.
We will explore models that incorporate each of these two limiting cases.

The second term on the righthand side of Equation (2) is the $z$-gradient
of the velocity dispersion cross-correlation, $\overline {v_R v_z}$. 
Binney and Tremaine (1987)
discuss two extreme possibilities, when the velocity dispersion
ellipsoid's principal axes remain aligned with the radial coordinate
of the galactocentric cylindrical
system, $R$, and when the ellipsoid's principal axes remain aligned
with the radial direction of the spherical system of coordinates.
In the first case, $\overline {v_R v_z}$ is independent of $z$,
and in the second case, the cross-term can be approximated as
$$
\overline {v_R v_z} \approx 0.5 \sigma_R^2{z \over R_{\odot}} \mbox{.}
$$
Taking into account that the vertical scale height of the Thick Disk is
smaller than its radial scale length, the cross-term in Equation (2)
can be approximated as:
$$
{R \over \rho_i \sigma_R^2}{\partial (\rho_i \overline {v_R v_z}) \over \partial z}
\approx 0.5 \Big( 1 - {z \over h_{z_{thick}}} \Big)
$$
The contribution to the disk rotation by the velocity dispersion cross-term
is small but non-negligible.
We shall model the equilibrium of the Thick Disk rotation with, and without the
cross-term. 

The remaining term within the parentheses in Equation (2), 
${\sigma_{\theta}}^2/{\sigma_R}^2$, is known from the proper-motion/photometry
analysis for our Thick Disk sample.
This leaves only the gravitational potential to be evaluated.

The gravitational potential, $\Phi_{tot}$, is the sum of the Disk 
potential and the potential of the Halo, 
$\Phi_{tot}(R,z) = \Phi_{disk}(R,z)+\Phi_{halo}(R,z)$.
The Disk potential is in turn a combination of the Thin-Disk gravitational 
potential and a contribution from the Thick Disk.
The Disk potential reflects the total mass of each disk component,
which is roughly proportional to the product of its density
in the plane with its vertical scale height.
Estimates of the density of the Galactic Thick Disk are rather uncertain,
with values of the thick-to-thin local density ratio
varying from about two percent (Gilmore 1984)
to about twenty percent (Fuhrmann 2004) with most of the recent estimates 
being in the range 3.5 to 7 percent
(Robin et al. 1996, Ojha et al. 2001, Du et al. 2003). 
On the other hand, with an ``exponential'' scale height of $\sim280$ kpc
(Korchagin et al. 2003),
the vertical scale of the Thin Disk is perhaps one third to one quarter 
that of the Thick Disk.
Thus, the total mass of the Thin Disk is significantly greater than that of
the Thick Disk.
Summarizing existing starcount studies, Siegel et al. (2002) find most
of these indicate the Thick Disk comprises roughly 10\% of the mass of
the Thin Disk. 
Furthermore, with the Thick Disk's presumed larger radial scale length, 
its contribution to the gradient of the total Galactic potential will be 
even less still.
For these reasons, we choose to neglect the Thick Disk's contribution to the
gravitational potential in studying its equilibrium.
As such, we need only specify models for the Thin Disk potential
and the Halo potential.

Making this explicit, and adopting expressions for the other terms as discussed
above, Equation (2) becomes
\be
{\overline {v_{\Theta}}}^2 (z) = \sigma_R^2 (z) 
\Big[ 
-{ \Upsilon_{a,b} R } +
0.5 \lambda \Big( 1 - {z \over h_{z_{thick}}} \Big) +
1 - {\sigma_{\Theta}^2 \over \sigma_R^2 }
\Big]
+  R {\partial \Phi_{halo} \over \partial R} +
R {\partial \Phi_{disk} \over \partial R}
\label{eq3}
\ee
where $\Upsilon_{a,b}$ assumes the form of either Equation (3a) or (3b) and
$\lambda$ is set to 1 or 0 depending on whether or not one chooses to
include the velocity-dispersion cross-term.
It now remains to adopt reasonable models for the Thin Disk
and Halo potentials.

\subsubsection{Thin Disk potential}
Assuming the surface-density distribution of the Thin Disk
to be exponential, $\Sigma_{thin} (R) = \Sigma_0 exp(-R/h_{R_{thin}})$, 
we can express
its gravitational potential in terms of the Bessel function (Toomre 1962),
\be
  \Phi_{disk} (R,z) = -2 \pi G \Sigma_0 h^2_{R_{thin}} \int^{\infty}_0 {J_0(kr) exp(-k|z|)k dk \over
[ 1 + k^2h_{R_{thin}}^2 ]^{3/2} }.
\label{eq5}
\ee
We define that portion of the Thick Disk rotation associated with the 
Thin-Disk gravitational potential to be $v_{disk}^2$, which can be expressed 
with the help of Equation (5) as
\be
v_{disk}^2 (R,z) \equiv R {\partial \Phi_{disk} \over \partial R} =
2 \pi G \Sigma_0 h^2_{R_{thin}} R_{\odot} \int^{\infty}_0 {J_1(kR) exp(-k|z|)k dk
\over [ 1 + k^2h_{R_{thin}}^2 ]^{3/2} }.
\label{eq6}
\ee
In the mid-plane of the disk, the integral in Equation (6) can be 
evaluated, and written in terms of the modified Bessel 
functions $I_i, K_i$ (Freeman 1970):
\be
v_{disk}^2(R_{\odot},0)  = 4 \pi G \Sigma_0 h_{R_{thin}} x^2
\Big[ I_0(x)K_0(x) - I_1(x) K_1(x) \Big ] \mbox{,}
\label{eq7}
\ee
where $ x = R_{\odot} / 2 h_{R_{thin}} $.

\subsubsection{Halo potential}
We consider the two simplest halo potential models, the Plummer model
(see Binney \& Tremaine 1987) and the pseudo-isothermal model 
(see Bahcall \& Soneira 1980).
The Plummer model potential is given by the expression
\be
\Phi_{halo} = - {G M_H \over (r^2 + a^2)^{1/2} } \mbox {,}
\label{eq8}
\ee
where $r$ is Galactocentric distance, $G$ is the gravitational constant, 
$M_H$ is the total mass of the halo, and $a$ is
the core radius.
The total halo mass can be expressed in terms of the rotational
velocity of the local standard of rest, $v_c$, which is determined
by the combined disk and halo gravitational potentials
in the solar neighborhood,
\be
v_c^2 = {G M_H R_{\odot} \over (R^2_{\odot} + a^2)^{3/2} } + 
v_{disk}^2(R_{\odot},0).
\label {eq9}
\ee
Solving for the mass of the halo, $M_H$, in Equation (9) and 
substituting it into Equation (4) yields
$$
{\overline {v_{\Theta}}}^2 (z) = \sigma_R^2 (z)
\Big[
-{ \Upsilon_{a,b} R } +
0.5 \lambda \Big( 1 - {z \over h_{z_{thick}}} \Big) +
1 - {\sigma_{\Theta}^2 \over \sigma_R^2 }
\Big]
$$
\be
+ { (v_c^2 - v^2_{disk}(R_{\odot},0))(R_{\odot}^2 + a^2)^{3/2}
\over (R_{\odot}^2 + z^2 +a^2)^{3/2} } +
v^2_{disk}(R_{\odot},z),
\label{eq10}
\ee
where the portion of the rotational velocity associated with 
the disk potential, both in and out of the plane,
$v^2_{disk}(R_{\odot},0)$ and $v^2_{disk}(R_{\odot},z)$, 
are given by Equations (6) and (7).

If, instead, a pseudo-isothermal model of the halo potential is assumed 
(Bahcall \& Soneira 1980),
expressions for the Thick-Disk rotational velocity can be derived
in a similar manner.
For this model, the halo potential is given by the expression
\be
\Phi_{halo} = 4 \pi G \rho_0 a^2 \Big[ 1 - {a \over r} \mbox{atan}{ r \over a} \Big] \mbox{,}
\label{eq11}
\ee
where $\rho_0$ is the central density of the halo.
The Thick-Disk rotational velocity then becomes
$$
{\overline {v_{\Theta}}}^2 (z) = \sigma_R^2 (z)
\Big[
-{ \Upsilon_{a,b} R } +
0.5 \lambda \Big( 1 - {z \over h_{z_{thick}}} \Big) +
1 - {\sigma_{\Theta}^2 \over \sigma_R^2 }
\Big]
+ 
$$
$$
\Big(
{  v_c^2 - v^2_{disk}(R_{\odot},0) \over  (a/R_{\odot}) \mbox{atan}(R_{\odot} / a ) -
(1 + R_{\odot}^2/a^2)^{-1} }\\
\Big)
\Big[ { a \over (R_{\odot}^2 +z^2)^{1/2} }
\mbox{atan} { (R_{\odot}^2 +z^2)^{1/2} \over a } - {1 \over  1 + (R_{\odot}^2 +z^2) / a^2 }\Big ]
$$
\be
\hspace{3cm}
+ v^2_{disk}(R_{\odot},z) \mbox{,}
\label{eq12}
\ee
where $v^2_{disk}(R_{\odot},0)$ and $v^2_{disk}(R_{\odot},z)$ are given 
by Equations (6) and (7).

\subsection{Results}

Equations (10) or (12), depending on one's choice of Halo potential, express
the equilibrium relationship between the Thick-Disk rotational velocity,
${\overline {v_{\Theta}}(z)}$, and the radial component of its
velocity dispersion, $\sigma_R(z)$.
We wish to test whether or not the intrinsic Thick Disk velocity and dispersion
profiles derived from our sample are consistent with these relations.
The profiles to be tested are those given in Section 3.4 by the parameters
($V|_{z=2.2}, V', \sigma_{U}|_{z=2.2}$, and $\sigma'_U$),
which are valid descriptions over the range $1 < Z < 4$ kpc.
As the relative uncertainties of the $V \equiv {\overline {v_{\Theta}}}$ 
profile are smaller than those of $\sigma_U \equiv \sigma_R$, we elect to
adopt the former, insert into Equations (10) and (12), and then
solve for the resultant dispersion profile.
The ``model'' dispersion profile can then be compared to that derived
from our measures, over the appropriate range in $z$.

In order to do so, various parameters of our equilibrium model must
be specified.
Values for these are taken from the literature as follows.

The Thin Disk contribution to the Galactic potential can be calculated
given its radial scale length, $h_{R_{thin}}$, and its mass 
surface density in the solar neighborhood, $\Sigma_{thin} (R_{sun})$. 
The Thin-Disk radial scale length is better determined than
that of the Thick Disk.
Recent estimates, (see e.g., Juri\'{c} et al. 2005), give a value
of 2.4 $\pm$ 0.2 kpc.
We adopt $h_{R_{thin}} = $2.5 kpc.
Korchagin et al. (2003) estimate the surface density of the Thin Disk in
the solar neighborhood to be 42 $\pm$ 6 M$_{\odot}$ pc$^{-2}$.
We shall explore values ranging from 36 to 48 M$_{\odot}$ pc$^{-2}$.
The value of $R_{sun}$ is taken to be 8 kpc.
 
The Thick-Disk radial scale length, $h_{R_{thick}}$, is one of the
key parameters that determines the rotational equilibrium
of the Thick Disk stars.
Estimates, based on star counts, 
range from 2.5 $\pm$ 0.3 kpc (Robin et al. 1996) 
to 4.3 $\pm$ 0.7 kpc (Larsen \& Humphreys 2003).
A recent determination based on the Sloan Digital Sky Survey data gives
a value of 3.5 $\pm$ 1 kpc (Juri\'{c} et al. 2005).
We will allow the value of $h_{R_{thick}}$ to vary from 3.5 to 5.0 kpc.

As mentioned in Section 4.1, we shall explore two different expressions
for the Thick-Disk pressure term, Equations (3a) and (3b).
These represent the limiting cases of behavior of the Thick Disk in terms
of the radial scale length of its velocity dispersion, i.e., does it
more resemble a self-gravitating disk, (3a), or a disk responding to
the external potential of the Thin Disk, (3b).

Another uncertainty is the inclusion of the velocity-dispersion cross-term,
regulated in Equations (10) and (12) by the factor $\lambda$.
We will generate models both with and without the cross-term included.
When included, the value for the Thick Disk's vertical scale height is
set to 0.785 kpc, as derived in Section 3.4.

Another of the terms that appears in Equations (10) and (12) is the
Thick-Disk velocity-dispersion ratio, $\sigma_{\Theta} /\sigma_R $.
Other studies find this ratio to be on the
order of one, or slightly less than one, in the solar neighborhood. 
(See for instance Beers \& Sommer-Larsen (1995) and Chiba \& Beers (2000).)
As can be seen from Figure 6, our measures agree and indicate that the ratio 
does not vary drastically as a function of $z$.
We will adopt our measures for this ratio, as given by the derived 
intrinsic dispersion profiles of Section 3.4.

The final model parameters to be specified are those related to the Halo.
A spherically symmetric halo is typically determined
by two parameters - the central density of the halo and its core radius.
Donato et al. (2004) studied the mass distribution in a sample of 25 
disk galaxies
of different morphological types using high-resolution rotation curves.
They find a strong correlation between the halo core radius, $a$, and
the thin-disk exponential scale length, $h_{R_{thin}}$, namely
$ a$[kpc] $\approx 13(h_{R_{thin}}$[kpc]$/5)^{1.05}$.
Using this relation, the Halo core radius of the Milky Way is
about 6.3 kpc.
We assume this value in our models. 
The central density of the dark matter Halo can be expressed via the
rotational velocity
of the local standard of rest, $v_c$, taking into account a contribution
from the Thin-Disk
gravitational potential given by Equation (9).
In our models, 220 km s$^{-1}$ 
is assumed for the velocity of the local standard
of rest.
And, of course, the choice of Equation (10) or Equation (12) reflects a
choice of the form of the Halo potential, i.e., that of a Plummer model or 
that of a pseudo-isothermal model.
Both will be examined.

Table 1 indicates the parameters used to calculate 32 equilibrium
models of the Thick Disk based on Equations (10) and (12) and our 
measured $V$-velocity profile from Section 3.4.
Each model generates a $U$-dispersion profile that is consistent
with the input $V$-velocity profile under the condition of equilibrium.
These calculated $U$-dispersion profiles are plotted in Figure 8, and
labeled by model number, as given in Table 1.

\begin{deluxetable}{ccccccccccccc}
\tabletypesize{\scriptsize}
\tablecaption{Equilibrium model parameters}
\tablewidth{0pt}
\tablehead{
\colhead{\tablenotemark{1}} &
\colhead{\tablenotemark{2}} &
\colhead{\tablenotemark{3}} &
\colhead{\tablenotemark{4}} &
\colhead{\tablenotemark{5}} &
\colhead{\tablenotemark{6}} &
&
\colhead{\tablenotemark{1}} &
\colhead{\tablenotemark{2}} &
\colhead{\tablenotemark{3}} &
\colhead{\tablenotemark{4}} &
\colhead{\tablenotemark{5}} &
\colhead{\tablenotemark{6}} \\
\colhead{Run} &
\colhead{$h_{R_{thick}}$} &
\colhead{$\Sigma_{thin}$} &
\colhead{Pl/} &
\colhead{$\lambda$} &
\colhead{a/b} &
\hspace{.5cm} &
\colhead{Run} & \colhead{$h_{R_{thick}}$} & \colhead{$\Sigma_{thin}$} &
\colhead{Pl/} & \colhead{$\lambda$} & \colhead{a/b} \\
\colhead{\#} & \colhead{$(kpc)$} & \colhead{$(M_{\sun}/pc^2)$} &
\colhead{iso} & \colhead{} & \colhead{} & &
\colhead{\#} & \colhead{$(kpc)$} & \colhead{$(M_{\sun}/pc^2)$} &
\colhead{iso} & \colhead{} & \colhead{}
}
\startdata
01 & 3.5 & 36.0 & Pl & 0 & a & &
                               05 & 3.5 & 36.0 & iso & 0 & a \\
02 & 5.0 & 36.0 & Pl & 0 & a & &
                               06 & 5.0 & 36.0 & iso & 0 & a \\
03 & 3.5 & 48.0 & Pl & 0 & a & &
                               07 & 3.5 & 48.0 & iso & 0 & a \\
04 & 5.0 & 48.0 & Pl & 0 & a & &
                               08 & 5.0 & 48.0 & iso & 0 & a \\
\tableline
09 & 3.5 & 36.0 & Pl & 1 & a & &
                               13 & 3.5 & 36.0 & iso & 1 & a \\
10 & 5.0 & 36.0 & Pl & 1 & a & &
                               14 & 5.0 & 36.0 & iso & 1 & a \\
11 & 3.5 & 48.0 & Pl & 1 & a & &
                               15 & 3.5 & 48.0 & iso & 1 & a \\
12 & 5.0 & 48.0 & Pl & 1 & a & &
                               16 & 5.0 & 48.0 & iso & 1 & a \\
\tableline
17 & 3.5 & 36.0 & Pl & 0 & b & &
                               21 & 3.5 & 36.0 & iso & 0 & b \\
18 & 5.0 & 36.0 & Pl & 0 & b & &
                               22 & 5.0 & 36.0 & iso & 0 & b \\
19 & 3.5 & 48.0 & Pl & 0 & b & &
                               23 & 3.5 & 48.0 & iso & 0 & b \\
20 & 5.0 & 48.0 & Pl & 0 & b & &
                               24 & 5.0 & 48.0 & iso & 0 & b \\
\tableline
25 & 3.5 & 36.0 & Pl & 1 & b & &
                               29 & 3.5 & 36.0 & iso & 1 & b \\
26 & 5.0 & 36.0 & Pl & 1 & b & &
                               30 & 5.0 & 36.0 & iso & 1 & b \\
27 & 3.5 & 48.0 & Pl & 1 & b & &
                               31 & 3.5 & 48.0 & iso & 1 & b \\
28 & 5.0 & 48.0 & Pl & 1 & b & &
                               32 & 5.0 & 48.0 & iso & 1 & b \\
\enddata
\tablenotetext{1}{
Model run number
}
\tablenotetext{2}{
Thick-Disk radial scale length
}
\tablenotetext{3}{
Local Thin-Disk mass surface density
}
\tablenotetext{4}{
Halo potential functional form; Plummer or isothermal
}
\tablenotetext{5}{
Velocity-dispersion cross-term factor
}
\tablenotetext{6}{
Form of the pressure term; Equation (3a) or (3b)
}
\end{deluxetable}

\begin{figure}
\epsscale{0.74}
\plotone{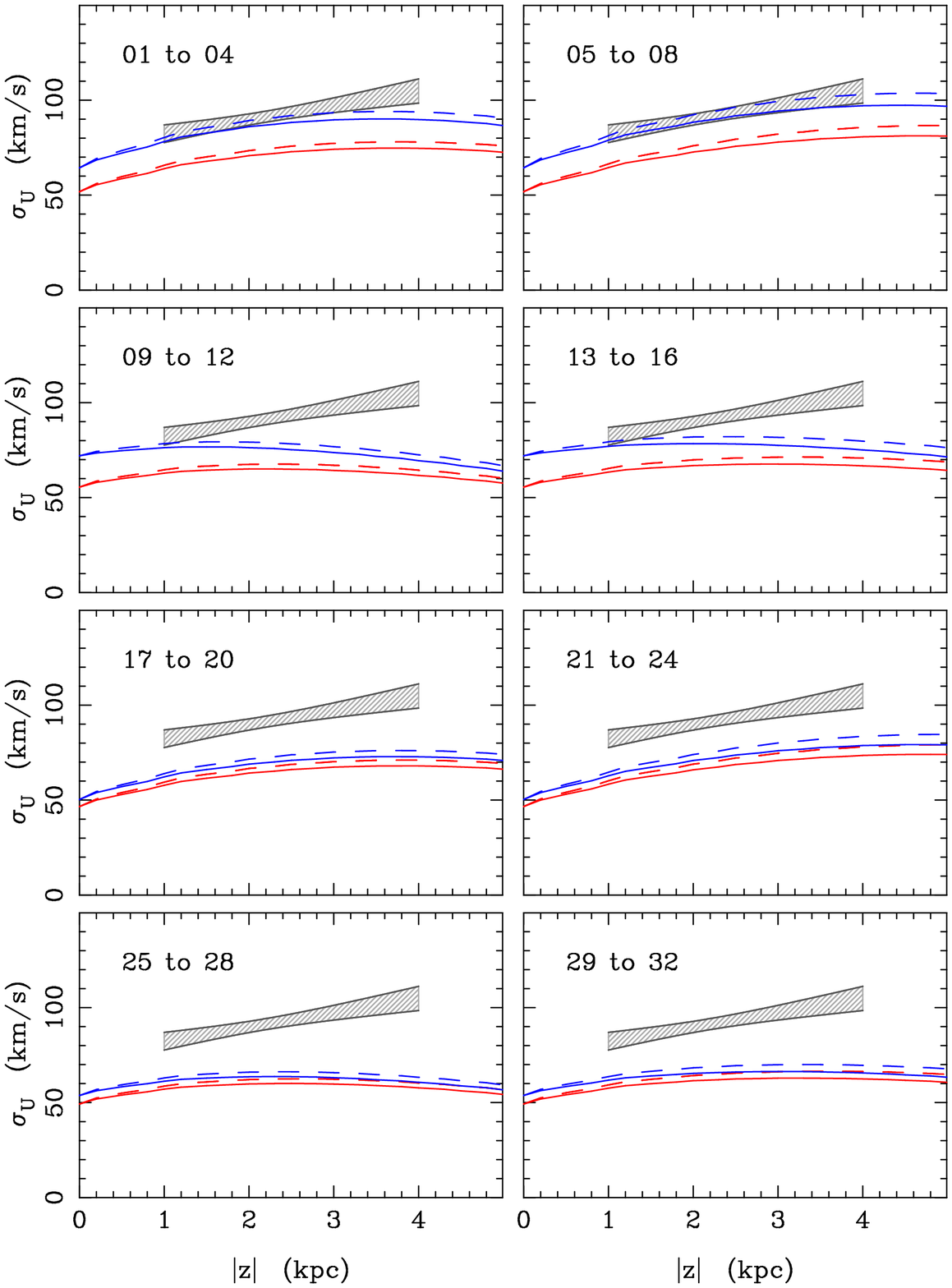}
\caption{
Velocity dispersion, $\sigma_U$, as a function of $z$, assuming our
equilibrium model and the $V$-velocity profile derived for our sample.
The 32 models shown correspond to those listed in Table 1 and explore
limiting values of five uncertain parameters of the equilibrium model.
Within each panel, the blue(/red) curves correspond to a value of the
Thick-Disk radial scale length of 5.0(/3.5) kpc.  The solid(/dashed)
curves are of models with the local surface density set to 48(/36)
M$_{\sun}$/pc$^2$.
The values of the remaining parameters can be inferred from the corresponding
model numbers in Table 1.
Also, within each panel is shown the $\pm$1-sigma error
range in the intrinsic $U$-disperion derived for our SGP sample.
}
\end{figure}

Within each panel of Figure 8 are four model profiles, corresponding to 
the upper and lower limits considered for the values of the Thick-Disk scale 
length and the Thin-Disk surface density.
Red curves are models with $h_{R_{thick}}$=3.5 kpc, while blue curves
are with $h_{R_{thick}}$=5.0 kpc.
Dashed curves are models with $\Sigma_{thin}(R_{sun})$=36 M$_{\sun}$/pc$^2$,
while solid curves have $\Sigma_{thin}(R_{sun})$=48 M$_{\sun}$/pc$^2$.
The column of four panels on the left represents a Plummer Halo potential
while the rightside column assumes a pseudo-isothermal form.
Models in the top four panels employ the ``self-gravitating'' form for
the pressure term, i.e., Equation (3a), while the bottom four panels are
of models using the ``external'' form, Equation (3b).
Finally, panels in the first and third rows are of models with no 
velocity-dispersion cross-term, while those in the second and fourth rows
include the cross-term.
Also shown in each panel, for the sake of comparison, is
the 1-sigma error range for the intrinsic $U$-dispersion profile
derived from the observed SGP proper-motion sample.

General trends seen in Figure 8 are worth noting.
The form of the pressure term, either Equation (3a) or (3b), affects
the $\sigma_U$ dispersion profile significantly.
The lower four panels, which are based on the assumption that the
Thick Disk sits within an ``external'' (Thin Disk) potential, reveal
consistently lower dispersion profiles.
Also, these models tend to be insensitive to the other Galactic parameters
such as the value of the local surface density, the radial scale length
of the Thick Disk, or the functional form of the Halo potential.
Conversely, in the top four panels, in which the pressure term is
that to be expected for a self-gravitating disk, the curves are sensitive
to the assumed value for the Thick-Disk radial scale length, and, to a
lesser extent, the form of the Halo potential.

In all cases, models that include the velocity-dispersion cross-term,
$\overline {v_R v_z}$, yield profiles that are relatively depressed,
either remaining level or decreasing slightly from $z$ = 1 to 4 kpc.
This is in contrast to the behavior of the derived intrinsic dispersion
profile based on our sample.

Judging from Figure 8, the Thick Disk velocity and 
velocity-dispersion profiles derived from our SGP sample are consistent
with one another and with the condition of equilibrium for at least some
subset of reasonable values of the various Galactic structure parameters.
In particular, the upper pair of curves in each of the two top panels
fit the data reasonably well.
These curves represent models with no velocity-dispersion cross-term,
with the ``self-gravitating'' form of the pressure term, and with the
higher estimate for the Thick-Disk radial scale length.
The pseudo-isothermal Halo model, the right panel, provides a better
fit to the data, although the Plummer-model fit is also acceptable.
The fits are largely insensitive to the value of the local surface density 
of the Thin Disk, i.e., dashed versus solid curves.

We note that
it is somewhat unsettling that the best fitting models are those that
assume the ``self-gravitating'' form of the pressure term for the Thick Disk.
Certainly a disk that has embedded within it another disk that is
substantially more massive would not expected to be self-gravitating,
and we think it unlikely that the Galactic Thick Disk is self-gravitating.
The underlying property of the models employing Equation (3a) is that
their pressure radial scale length is substantially larger than the mass scale
length of the Thin Disk, approximately by a factor of two.
Rather than assuming a direct link to the radial scale length of the 
Thick Disk's mass distribution and self-gravitation,
this might also represent a failure of a second assumption
inherent in Equation (3a), that the shape of the Thick-Disk velocity ellipsoid
remains unchanging as a funtion of $R$ in the local neighborhood.
This, in turn, might be due to scattering of Thick Disk stars that is 
non-isotropic, perhaps by dense spiral arms.
This is purely speculative, though.
Further proper-motion studies at low
Galactic latitudes, combined with radial-velocity measures, would help
address the nature of the Thick Disk's velocity ellipsoid and its spatial
variation.

\section{Discussion}

\subsection{Comparison to Previous Studies}

Our estimate of the Thick-Disk scale height, $783 \pm 48$ pc, is
consistent with previous determinations from other star-count studies,
albeit possibly on the low end of said range.
Buser et al. (1999), in their Table 5, and Siegel et al. (2002), 
in their Table 1, provide summaries of previous estimates of the Thick-Disk
scale height.
Values range from 700 to 2000 pc, with most estimates falling between 800
and 1200 pc.
More recently, a study by Cabrera-Lavers et al. (2005), identifying 
Red Clump stars from 2MASS data, yields a value of $1062 \pm 52$ pc.
Using SDSS data, Juri\'{c} et al. (2005) find a value of 1200 pc, 
based on dwarf stars within 1.5 kpc of the Sun.
Siegel et al. (2002) point out the inverse
correlation between derived scale height and
local density normalization in some Thick-Disk modeling studies.
Because of the manner in which our SGP sample was selected, we do not
attempt to estimate the local density of the Thick Disk component.

The rotational lag of the Thick Disk has been measured in a number of studies.
Majewski (1993) summarizes those made prior to 1993 in his Figure 7 of that
review.
These estimates vary substantially, from roughly -20 km s$^{-1}$ to 
-120 km s$^{-1}$,
and with a suggestion of a correlation with $z$, the larger lag values being
found for samples at larger $z$.
Several of the studies that Majewski references fall nicely along our
relation of $z$ and Thick Disk lag.
Among these are the studies of Norris (1986), Wyse \& Gilmore (1990), 
and Beers et al. (1992).
Several other studies, specifically those of Murray (1986) and of Hanson (1989)
obtain values for the $z$ gradient in lag that are similar to what we find.
While not all of these $z$-gradient studies
isolate the Thick Disk stars in their samples,
as Majewski (1993) argues, it is unlikely that Halo contamination can
account entirely for the observed gradient.
In Majewski's (1992) proper-motion study of the North Galactic Pole (NGP), the
Thick Disk is fit through multi-component modeling.
He finds a somewhat lower gradient in the lag, 
-21 $\pm$ 1 km s$^{-1}$ kpc$^{-1}$.
More recently, in a full space-motion study by Chiba \& Beers (2000),
in which the Thick Disk component is separated by metallicity, they
obtain the same value for the gradient as do we, 
-30 $\pm$ 3 km s$^{-1}$ kpc$^{-1}$.
In another 3-d velocity study, at the NGP, 
Soubiran et al. (2003)
do detect a separate Thick Disk component, via the kinematics, but do not
see overwhelming evidence for a $z$ gradient, although this is not ruled out.
Instead they present a single value of -51 $\pm$ 5 km s$^{-1}$
as representative of their Thick Disk sample.
Gilmore et al. (2002) analyze
radial velocities at intermediate-latitude lines-of-sight along
Galactic rotation and anti-rotation.
Their model of the velocity distribution fits the observations much better
when their fainter, more distant sample is modeled using a Thick
Disk with a substantial rotational lag, i.e., -100 km s$^{-1}$.
Wyse et al. (2006), further examine the samples along these lines-of-sight
and again see evidence of a high-lag component, 
$\sim$ -120 km s$^{-1}$, in
addition to a ``canonical'' Thick Disk with a constant lag of -40 km s$^{-1}$.
We note, however, that the excess they observe might very well be explained
by a Thick Disk component with substantial lag and $z$-gradient, such
as our SGP sample implies.

It does not seem likely that Halo contamination can account for the large
$z$-gradient that we observe in the lag.
Our sample's number-density profile provides an estimate of the 
Halo contamination, a modest 8 $\pm$ 6 \%, and this has been included in
our simulated samples from which the intrinsic Thick Disk lag was derived.
Nor is it likely that random errors in the proper motions or
in the distance estimates would lead us to 
overestimate the lag and its gradient.
These uncertainties have very little effect on the observed $V$ profile,
because of the geometry involved and the simple relationship between distance,
proper motion, and velocity.
A systematic error in the absolute proper-motions would translate directly
into a systematic error in the $V$-velocity gradient.
The estimated uncertainty in the proper-motion system of the SPM3
is 0.4 mas yr$^{-1}$, corresponding to an error of 
only 2 km s$^{-1}$ kpc$^{-1}$ in the $z$-gradient
of either of the tangential-velocity profiles.

Such a systematic error could account for the marginally significant
gradient found in the $U$ velocity, 4.8 $\pm$ 2.9 km s$^{-1}$ kpc$^{-1}$.
A flat $U$-velocity profile is consistent with disklike motion
for the sample.
The offset of 19.1 $\pm$ 2.7 km s$^{-1}$ is presumably a reflex 
of the peculiar motion of the sun in this direction.
Although, this value is larger than expected when compared to the
Hipparcos-based value of 10.0 $\pm$ 0.4 determined by Dehnen \& Binney (1998). 

While the velocity profiles are sensitive to systematic errors, it is the
random errors in both proper motion and distance estimates that affect the
observed velocity-dispersion profiles.
Figure 9 shows the resultant ``observed'' $U$-dispersion profile of our
best-fit model, in comparison to the underlying intrinsic profile.
The profile is significantly steepened, illustrating the need for
accurate knowledge of the observational uncertainties when attempting to
discern the intrinsic velocity dispersion.

\begin{figure}
\epsscale{0.80}
\plotone{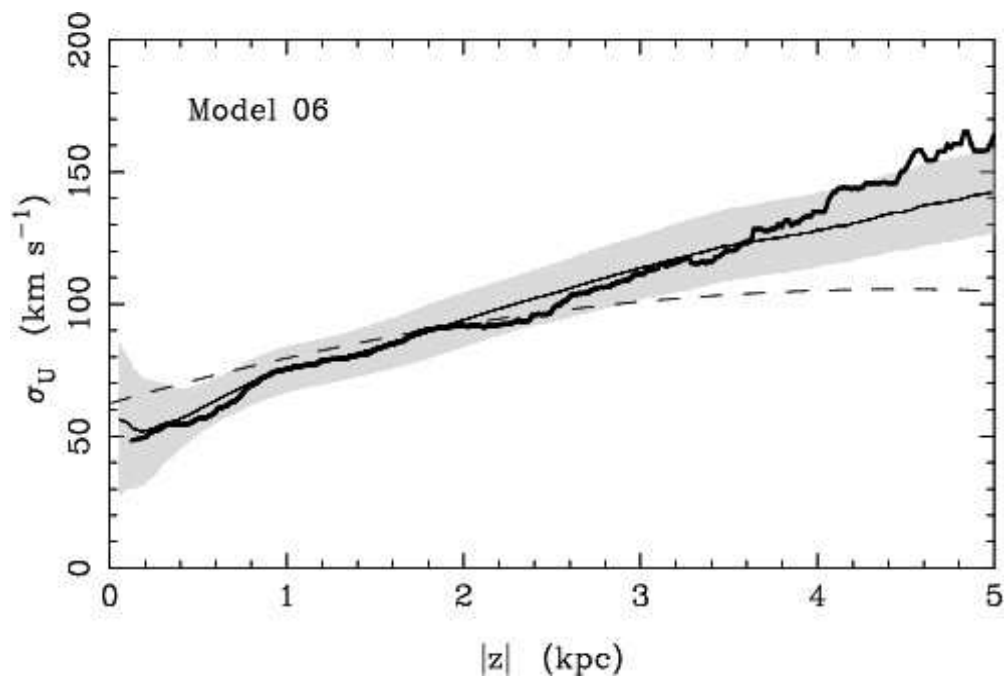}
\caption{
Intrinsic and ``observed'' velocity dispersions, $\sigma_U$, as a function of
$z$, for the best-fit equilibrium model, \#06.
The dashed curve shows the intrinsic dispersion profile that results from
this model.
The thin solid curve, with $\pm$1-sigma error bars shown in gray, indicates
the mean of 100 Monte-Carlo simulated datasets based on this intrinsic
dispersion profile.
The heavier, irregular line shows the $\sigma_U$ profile observed for the
actual SGP red-giant sample.
}
\end{figure}

With this in mind, the velocity dispersions we derive
show a modest $z$-gradient of about +9 km s$^{-1}$ kpc$^{-1}$ 
over the range $1<z<4$.
At our sample's mean distance of $z$=2.2 kpc, the dispersions are
roughly 90 and 70 km s$^{-1}$ in $\sigma_U$ and $\sigma_V$, respectively.
This can be compared to the velocity dispersion of the intermediate-metallicity
sample of Chiba \& Beers (2000), which they identify with the Thick Disk.
They find $\sigma_V$ increases from roughly 40 km s$^{-1}$ near the plane, 
to a value of about 70 km s$^{-1}$ at $z$ = 0.5 kpc, remaining constant 
after that out to their sample's limiting distance of $z$ = 1.75 kpc.
Majewski (1994) has summarized the results of six different proper-motion
studies toward the Galactic Poles.
He shows in his Figure 1 the $z$ dependence of $\sigma_U$ and $\sigma_V$ from
these studies.
It is important to note that these samples include {\it all} stars, with no
attempt to isolate Thick Disk members, although over some range of $z$ the
Thick Disk should dominate naturally.
The various studies agree reasonably well, showing a linear increase in
$\sigma_U$ and $\sigma_V$ from 10 to 20 km s$^{-1}$ in the solar neighborhood
to approximately 70 to 100 km s$^{-1}$ at $z$ = 2 kpc.
This steep gradient is presumably due to the Thin Disk stars' lower
velocity dispersion in the solar neighborhood.
On the other hand, the general agreement with our derived dispersions near
$z$ = 2 kpc suggests that the Thick Disk dominates all samples at this range.
In Majewski's (1992) study, where an effort $is$ made to isolate the
Thick Disk stars, a gradient of $\sim$ 12 km s$^{-1}$ kpc$^{-1}$ is found 
out to $z$ = 4 kpc, similar to what we obtain.

\subsection{Implied Nature of the Thick Disk}

The original notion of thick disks was introduced in the surface-photometry
studies of S0 galaxies by Burstein (1979) and Tsikoudi (1979).
Subsequent observations, both Galactic and extragalactic, have begun to 
provide an understanding of the nature and possible formation processes
of thick disks. 
The current picture 
for disk galaxies is detailed in Yoachim \& Dalcanton (2006) and
references therein. 
In short, they find that thick disks are a somewhat
regular component of disk galaxies, that the scale heights and scale lengths
of thick disks exceed those of their embedded thin disks by factors
of 1.25 and 2, respectively, and that they are made up of
old, metal-poor stars, (see also Mould et al. 2005, Davidge
2005, Tikhonov et al. 2005).  
Yoachim \& Dalcanton also find that, for less
massive galaxies, the thick disk contributes significantly
(up to half) to the luminous mass. 
From the kinematical analysis of two different
disk galaxies, Yoachim \& Dalcanton (2005) find that the
thick disk rotates at $30-40\%$ of the rotation speed of the thin disk
in one of the galaxies, while
in the second, the thick disk seems to counterrotate, if it rotates at all.

With regard to the Thick Disk of our own Galaxy,
there is mounting evidence that the Thick Disk's
chemical abundance profile is distinct from that of the 
Thin Disk, (for a more thorough discussion, see 
Brewer \& Carney 2006 and references therein).
Specifically, Thick Disk stars have a ratio $[\alpha/$Fe] for given Fe
that is systematically higher than the $[\alpha/$Fe] ratio for Thin Disk stars. 
Also, there is little or no
gradient in metallicity as a function of distance from the Galactic plane,
(Gilmore et al. 1995, Rong et al. 2001). 
These properties point to the Thick Disk
being a system that is distinct from the present Thin Disk.

There are a large number of studies that have attempted to model
the formation of the Galactic Thick Disk. 
Briefly, these can be classified into three main groups: 
1) the Thick Disk is the precursor of the Thin Disk,
and was formed from gas at a large scale height that settled 
in a slow, pressure-supported collapse
(see e.g., Gilmore \& Wyse 1986, Burkert et al. 1992),
2) the Thick Disk formed from a pre-existing Thin Disk by a
heating mechanism, the most viable of which is energy deposited by the
accretion of a dense
satellite (see e.g., Quinn et al. 1993, Velazquez \& White 1999),
although other mechanisms have been proposed (Kroupa 2002), and 
3) the Thick Disk is formed from the accretion and disintegration
of satellites
during an early and active period of hierarchical clustering
(Abadi et al. 2003, Brook et al. 2004).
This last scenario relies on satellites having orbits highly
coupled with the disk rotation, i.e., orbits of high circularity
and/or that are strongly circularized due to dynamical friction.

The evidence from chemical abundances of Thick Disk stars appears to
favor either of the last two scenarios. 
Further distinguishing between these last 
two types of formation might be possible by comparing the
observed kinematical profiles of the Thick Disk to formation models.
Among recent simulations, there are regrettably few that present
velocity-lag and velocity-dispersion profiles. 
Instead, these works tend to emphasize the density profile, 
chemical abundance pattern and 
vertical velocity dispersion as a function of age,
and generally present simple means for the tangential velocity dispersions. 
A meaningful comparison of our results to theoretical models is therefore
difficult.
However, we mention two examples that might serve as starting points 
for more comprehensive
kinematical comparisons between formation models and observations.

The study of Abadi et al. (2003) does present a prediction
for the rotational lag of the combined disks as a function of $z$. 
Interestingly, their
Figure 5 indicates a gradient of $\sim -20$ km s$^{-1}$ kpc$^{-1}$ in the range 
$2 < z < 4$ kpc where the thick disk dominates. 
In the Abadi et al. simulations, well over half of the thick-disk stars are
the debris from disrupted satellites, circularized through dynamical friction
before being absorbed.

The second example we note is that of Velazquez \& White (1999). 
This study analyzes the thickening of
a pre-existing thin disk due to the accretion of a single, dense, 
massive (20\% of the mass of the disk) satellite.
While not presenting the resulting ``thick-disk'' velocities or 
velocity dispersions as a function of $z$, they do show the $R$ profiles
of the velocity dispersions.
In particular, from their Figure 5, the $R$-gradient of $\sigma_U$ for
the resulting thick disk is slightly shallower than that of the
precursor thin disk.
We note that qualitatively this is in agreement with our equilbrium
modeling, for which the best fits were found using the form of the
pressure term that corresponds to a shallower gradient, our Equation (3a).

Although we are unable to distinguish between current scenarios for
Thick Disk formation at this time, we urge those who construct such
models to present the resulting $z$ profiles of the various velocity
components and dispersions.
The observational constraints provided by proper-motion studies such
as this one, as well as radial-velocity studies, will only continue to
improve with time.

\section{Summary}

SPM3 absolute proper motions are combined with 2MASS near-infrared
photometry for a photometrically-selected sample of $\sim$1200 stars at the
South Galactic Pole.
It is shown that these stars are predominantly Thick-Disk red giants.
Photometric distances and tangential velocity distributions are
determined for the sample using a Monte-Carlo modeling method that
automatically corrects for bias due to distance errors.
From this modeling, the intrinsic properties of the Thick Disk are derived
as a function of distance from the Galactic plane over the range $1<z<4$ kpc.

The vertical scale height of the Thick Disk is found to be
783 $\pm$ 48 pc.
The rotational lag shows a linear variation with $z$, having a gradient
of -30.3 $\pm$ 3.2 km s$^{-1}$ kpc$^{-1}$ 
and a value of -91.6 $\pm$ 1.9 km s$^{-1}$ at the
samples' mean distance of 2.2 kpc.
The tangential velocity dispersions also show a variation with $z$, although
with a much shallower slope of about 9 $\pm$ 3 km s$^{-1}$.
Under the assumption of dynamical equilibrium, the $V$-velocity profile and
(primarily) $U$-velocity dispersion profile as functions of $z$ are
constrained by the Galactic potential within which the Thick Disk resides.
We have shown that for reasonable parameters of the dominant Galactic
components, i.e., the Halo and Thin Disk, the kinematics of the Thick
Disk as derived from our observed sample are self-consistent.
Our modeling also indicates preference for a Thick-Disk 
velocity ellipsoid with zero cross term, 
i.e., one whose axes remain parallel to the Galactic plane independent
of $z$.
The tangential velocity and velocity-dispersion profiles presented here
should be able to better constrain theoretical models of Thick Disk formation.

We are grateful to an anonymous referee for making a number of helpful
comments, among them, the suggestion to include 
the reduced proper motion diagram.
van Altena wishes to thank Dr. M. Miyamoto and Dr. M. Yoshizawa for the
opportunity to spend a sabbatical semester, back in 1995, at the National
Astronomical Observatory in Mitaka, Japan.
Discussions with them and co-author Korchagin were important to initiating
this research.
This publication makes use of data products from the Two Micron All Sky Survey.
The Yale/San Juan SPM program has been supported over the years by a number
of grants from the National Science Foundation, the most recent of these
being NSF grants 0098548 and 0407293.


\begin{references}{}

\reference{}
Abadi, M. G., Navarro, J. F., Steinmetz, M., \& Eke, V. R. 2003, ApJ, 597, 21

\reference{}
Allende Prieto, C., Beers, T. C., Wilhelm, R., Newberg, H. J., Rockosi, C. M.,
Yanny, B., Lee, Y. S. 2006, ApJ, 636, 804

\reference{}
Bahcall, J. N., \& Soneira, R. M. 1980, ApJS, 44, 73

\reference{}
Beers, T. C., Doinidis, S. P., Griffin, K. E., Preston, G. W., Shectman, S. A.
1992, AJ, 103, 267

\reference{}
Beers, T. C. \& Sommer-Larsen, J. 1995, ApJS, 96, 175

\reference{}
Binney, J., \& Tremaine, S. 1987, Galactic Dynamics, 
(Princeton: Princeton Univ. Press)

\reference{}
Bland-Hawthorn, J., \& Freeman, K. 2004, PASA, 21, 110

\reference{}
Brewer, M. M., \& Carney, B. W. 2004, PASA, 21, 134

\reference{}
Brewer, M. M., \& Carney, B. W. 2006, AJ, 131, 431

\reference{}
Brook, C. B., Kawata, D., Gibson, B. K., \& Freeman, K. C. 2004, ApJ, 612, 894

\reference{}
Burkert, A., Truran, J. W., \& Hensler, G. 1992, ApJ, 391, 651

\reference{}
Burstein, D. 1979, ApJ, 234, 829

\reference{}
Buser, R., Rong, J., \& Karaali, S. 1999, A\&A, 348, 98

\reference{}
Cabrera-Lavers, A., Garz\'{o}n, F., \& Hammersley, P. L. 2005, A\&A, 433, 173

\reference{}
Carney, B. W., Latham, D. W., \& Laird, J. B. 1989, AJ, 97, 423

\reference{}
Chen, B., Stonghton, C., Smith, J. A., et al. 2001, ApJ, 553, 184

\reference{}
Chiba, M., \& Beers, T. C. 2000, AJ, 119, 2843

\reference{}
Dehnen, W. \& Binney, J. J. 1998, MNRAS, 298, 387

\reference{}
Donato, F., Gentile, G., \& Salucci, P. 2004, MNRAS, 353, L17

\reference{}
Du, C., Zhou, Hu, Ma, Jun, et al. 2003, A\&A, 407, 541
 
\reference{}
Eggen, O. J., Lynden-Bell, D., \& Sandage, A.R. 1962, ApJ, 136, 748

\reference{}
Freeman, K. C. 1970, ApJ, 160, 811

\reference{}
Fuhrmann, K. 1998, A\&A, 338, 161

\reference{}
Fuhrmann, K. 2004, Astron.Nachr., 325, 3

\reference{}
Gilmore, G., \& Wyse, R. F. G., \& Norris, J. E. 2002, ApJ, 574, L39

\reference{}
Gilmore, G. 1984, MNRAS, 207, 223

\reference{}
Gilmore, G., \& Wyse, R. F. G. 1986, Nature, 322, 806

\reference{} 
Gilmore, G., Wyse, R. F. G., \& Jones, J. B. 1995, AJ, 109, 1095

\reference{}
Gilmore, G., \& Reid, N. 1983, MNRAS, 202, 1025

\reference{}
Girard, T. M., Dinescu, D. I., van Altena, W. F., Platais, I., Monet, D. G.,
\& L\'{o}pez, C. E. 2004, AJ 127, 3060

\reference{}
Hamaker, H. C. 1978, Applied Statistics, 27, 76

\reference{}
Hanson, R. B. 1989, BAAS, 21, 1107

\reference{}
Holmberg, J., \& Flynn, C. 2000, MNRAS, 313, 209

\reference{}
Jackson, T., Ivezi\'{c}, Z., \& Knapp, G. R. 2002, MNRAS, 337, 749

\reference{}
Juri\'{c}, M., Ivezi\'{c}, Z., Brooks, A., et al. 2005, astro-ph/0510520

\reference{}
Korchagin, V. I., Girard, T. M., Borkova, T. V., Dinescu, D. I., \&
van Altena, W. F. 2003, AJ 126, 2896

\reference{}
Kroupa, P. 2002, MNRAS, 330, 707

\reference{}
Larsen, J. A., \& Humphreys, R. M. 2003, ApJ, 125, 1958

\reference{}
Majewski, S. R. 1992, ApJS, 78, 87

\reference{}
Majewski, S. R. 1993, Ann. Rev. Astron. Astrophys. 31, 575

\reference{}
Majewski, S. R. 1994, in Astronomy from wide-field imaging: proceedings of 
IAU Symp. 161,
ed. H. T. McGillivray, E. B. Thomson, B. M. Lasker, I. N. Reid, D. F. Malin,
R. M. West \& H. Lorenz (Dordrecht: Kluwer), 425

\reference{}
Murray, C. A. 1986, MNRAS, 223, 649

\reference{}
Norris, J. E. 1986, ApJS, 61, 667

\reference{}
Norris, J. E. 1999, Ap\&SS, 265, 213

\reference{}
Ojha, D. K. 2001, MNRAS, 322, 426

\reference{}
Parker, J. E., Humphreys, R. M., \& Beers, T. C. 2004, AJ, 127, 1567

\reference{}
Prohaska, J. X., Naumov, S.O., Carney, B. W., McWilliam, A., \& Wolfe, A. M.
2000, AJ, 120, 2513

\reference{}
Quinn, P. J., Hernquist, L., \& Fullagar, D. P. 1993, ApJ, 403, 74

\reference{}
Reid, N., \& Majewski, S. R. 1993, ApJ, 409, 635

\reference{}
Robin, A.C., Haywood, M., Creze, M., et al. 1996, A\&A, 305, 125

\reference{}
Rong, J., Buser, R., \& Karaali, S. 2001, A\&A, 365, 431

\reference{}
Siegel, M. H., Majewski, S. R., Reid, I. N., \& Thompson, I. B. 2002, 
AJ, 578, 151

\reference{}
Soubiran, C., Bienaym\'{e}, O., \& Siebert, A. 2003, A\&A, 398, 141

\reference{}
Toomre, A. 1962, ApJ, 138, 385

\reference{}
Tsikoudi, V. 1979, ApJ, 234, 842

\reference{}
Tsuchiya, T., Korchagin, V. I., \& Dinescu, D. I. 2004, MNRAS, 350, 1141

\reference{}
Velazquez, H. \& White, S. D. M. 1999, MNRAS, 304, 254

\reference{}
Wyse, R. F. G. \& Gilmore, G. F. 1990, in {\it Chemical \& Dynamical 
Evolution of Galaxies}, eds. F. Ferrini, J. Franco, F. Matteucci, p. 19,
(Pisa:ETS Editrice)

\reference{}
Wyse, R. F. G. \& Gilmore, G. 1995, AJ 110, 2771

\reference{}
Wyse, R. F. G., Gilmore, G. Norris, J. E., Wilkinson, M. I., Kleyna, J. T.,
Koch, A., Evans, N. W., \& Grebel, E. K. 2006, ApJL, 639, 13

\reference{}
Yi, S. K., Kim, Y.-C., \& Demarque, P. 2003, ApJS 144, 259

\reference{}
Yoachim, P. \& Dalcanton, J. 2005, ApJ, 624, 701

\reference{}
Yoachim, P. \& Dalcanton, J. 2006, AJ, 131, 226

\reference{}
Zinn, R. 1985, ApJ, 293, 424

\end{references}
\end{document}